\documentclass[12pt]{article}
\usepackage[italian,english]{babel}
\usepackage{graphicx}
\usepackage{verbatim}
\usepackage{epsf}
\usepackage{latexsym}
\usepackage{amsmath,amsfonts,amssymb,amsthm}
\def\bseq{\begin{subequation}}  
\def\eseq{\end{subequation}}
\def\bsea{\begin{subeqnarray}}  
\def\esea{\end{subeqnarray}}

\newcommand{\bbox}{\lower.2ex\hbox{$\Box$}}

\newcommand{\beq}{\begin{equation}}
\newcommand{\eeq}{\end{equation}}
\newcommand{\bea}{\begin{eqnarray}}
\newcommand{\eea}{\end{eqnarray}}
\newcommand{\ena}{\end{eqnarray}}

\newcommand {\non}{\nonumber}

\newcommand{\Tr}{{\rm Tr}}

\newcommand{\be}{\begin{equation}}
\newcommand{\ee}{\end{equation}}

\begin{document}
\setcounter{page}{0}
\begin{titlepage}
\titlepage
\begin{flushright}
CERN-PH-TH/2008-039\\
SISSA 10/2008/EP\\
\end{flushright}
\vskip 3cm
\centerline{{ \bf \Large Metastable vacua and geometric deformations}}
\vskip 0.5cm
\centerline{{ \bf \Large  }}
\vskip 1.5cm
\centerline{Antonio Amariti$^{a}$, Davide Forcella$^{b}$, Luciano
Girardello$^{a}$, Alberto Mariotti$^{a,c}$}
\begin{center}
$^a$ Dipartimento di Fisica, Universit\`a di Milano Bicocca\\
and \\
INFN, Sezione di Milano-Bicocca,\\
piazza della Scienza 3, I-20126 Milano, Italy
\vskip .4cm
$^b$ International School for Advanced Studies (SISSA/ISAS) \\
and \\
INFN-Sezione di Trieste,\\
via Beirut 2, I-34014, Trieste, Italy\\
\vskip .4cm
$^b$ PH-TH Division, CERN CH-1211 Geneva 23, Switzerland\\
\vskip .4cm
$^c$
Service de Physique Theorique, SPhT \\
Orme des Merisiers, CEA/Saclay \\
91191 Gif-sur-Yvette Cedex, FRANCE \\
\vskip .4cm
$^c$
LPTHE, Universites Paris VI, Jussieu\\
F-75252 Paris, FRANCE\\
\end{center}
\vskip 1.5cm
\begin{abstract}
  We study the geometric interpretation of metastable vacua for
  systems of D3 branes at non isolated toric deformable singularities.
  Using the $L^{aba}$ examples, we investigate the relations between
  the field theoretic susy breaking and restoration and the complex
  deformations of the CY singularities.
\end{abstract}

\vfill
\begin{flushleft}
{\today}\\
\end{flushleft}
\end{titlepage}

\newpage

\tableofcontents

\addcontentsline{toc}{section}{Introduction}
\section*{Introduction}

The ISS mechanism \cite{Intriligator:2006dd}, based on long living
metastable vacua, greatly increases the class of gauge theories with
chiral matter and dynamically broken supersymmetry.  Much work has
indeed followed, in different directions
\cite{Franco:2006es}.

It has prompted a search for a string approach: either within the
gauge/gravity correspondence or toward a more direct string origin or
interpretation \cite{Franco:2006ht}-\cite{Marsano:2008ts}. These
remain open problems and only partial results are at hand.

Recently, some steps have emerged for the grounding of a geometrical
interpretation of the features of metastability in simple quiver
gauge theories
on $D$-branes near a singularity inside a CY manifold
\cite{Buican:2007is,Aganagic:2007py}.
The aim is to phrase the metastable $F$-type susy breaking in
a general geometrical language.
A key point is that the non perturbative dynamics behind the existence
of metastable vacua corresponds to deformations of a theory with
unbroken suspersymmetry \cite{Buican:2007is}.
The deformations regard the superpotential:
in the $D$-brane setting of IIB string theory they are mapped
into complex deformations in the local geometry.

In this paper we develop this approach further. We
study systems of branes at toric conical Calabi-Yau singularities of a
special type, i.e. deformable singularities, in the sense of Altman's
deformations \cite{Altman}, that are not isolated.  These form a large
subfamily of toric singularities and consist of a cone with a
singularity at the tip and some set of lines of
$\mathbb{C}^2/\mathbb{Z}_n$ singularities passing through it.
Different combinations of fractional branes at these singularities
give rise to different IR behaviors of the gauge theory:
$\mathcal{N}=2$ dynamics, confinement, runaway supersymmetry breaking
\cite{Franco:2005zu}, and long living metastable vacua, as recently
pointed out in \cite{Buican:2007is}.  Some of the different IR
dynamics can be geometrically understood as motion in the moduli space
of the CY singularities.

Our discussion is mainly focused on metastability in the
quiver gauge theories living on deformed $L^{aba}$
singularities.
Such theories correspond to an infinite class of non isolated
toric singularities, with a known metric.
Beyond their role in model building and in the
gauge/gravity duality, they form a fitting laboratory for
the investigation of the field theory/geometry correspondence.
In the analysis of general $L^{aba}$ quivers we show that
we can always extract subclasses
where metastable vacua exist.
The features of broken and restored supersymmetry find a systematic
geometric counterpart in terms of appropriate deformations of the geometry
of the unbroken susy phase.

The plan of the paper is as follows.
In section \ref{SPPsing} we review the case of the Suspended
Pinch Point ($SPP$) singularity, its associated field theory and
the relation between their corresponding deformations. This simple
case will be the guideline for the whole paper.
In section \ref{labasing} we
introduce the family of $L^{aba}$ singularities and the corresponding
quiver gauge theories.  We then analyze the metastable vacua in the
$L^{aba}$ gauge theories with $b \neq a$ in section \ref{metlaba}, and
the $L^{aaa}$ gauge theories
in section \ref{metlaaa}.
In all these cases we show that some deformation of the geometry leads
to metastability and some other deformation restores supersymmetry.
Metastability turns out to be a quite generic phenomenon in these
deformed toric theories.
Finally, in section \ref{metgen} we try to extend this
analysis to more complicated singularities.
Since we shall use some elements of toric geometry we present
in Appendix \ref{APPTORICA} a lightening review of a
few aspects and instructions for drawing out information
of interest in our investigation.
In the Appendix \ref{ISSmodello} we review the ISS
model and discuss the issue of gauging flavour.  In the Appendix
\ref{secagana} we outline the technique introduced in
\cite{Aganagic:2007py} for the computation of the superpotential from
the geometry.  In the Appendix \ref{appB} we give details on the non
supersymmetric vacua analyzed in the paper.  In the Appendix
\ref{UVcomplete} we discuss the problem of UV completion in a
clarifying example.

\section{Complex deformations and metastability: the SPP example}
\label{SPPsing}
\begin{figure}[h!!!]
\begin{center}
\includegraphics[scale=0.55]{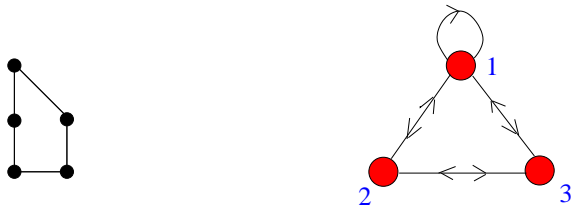}
\caption{The toric diagram and the quiver of the $SPP$ singularity}
\label{tqspp}
\end{center}
\end{figure}

The SPP gauge theory \cite{Douglas:1997de} is
obtained as the near horizon limit
of a stack of D3 branes on the tip of the conical singularity
\begin{equation}\label{spp}
x y^2 = w z  \, .
\end{equation}
The holomorphic equation defining the singularity can be encoded in a
graph called the toric diagram (see the Appendix \ref{APPTORICA}).
In the paper we will use
these diagrams to give an intuitive visual picture of the
singularities.
\\
The field theory has $U(N_1)\times U(N_2) \times U(N_3)$ gauge groups and
chiral superfields
that transform in the adjoint and bifundamental
representations of the various gauge group factors. The fields $X_{ii}$ are in
the adjoint of the i-th gauge group and the fields $q_{ij}$ transform in the
fundamental representation of the $U(N_i)$ gauge group and in the
anti-fundamental representation of the $U(N_j)$ gauge group.
The symmetries and the matter content of a gauge theory
related to branes at singularities can be encoded
in a graph called the quiver diagram.
The toric diagram and the quiver of the $SPP$ singularity
are shown in Figure \ref{tqspp}.
\\
\\
Its superpotential is\footnote{The superpotential is a sum of gauge
  field monomials obtained contracting gauge indexes and taking the
  trace. Explicit index contractions and traces will be omitted in the
  paper.}
\begin{equation}\label{superpotspp}
W= X_{11}(q_{13}q_{31}- q_{12}q_{21} ) + q_{21}q_{12}q_{23}q_{32} -
q_{32}q_{23}q_{31}q_{13} \, .
\end{equation}
Taking into account the F-term equations for (\ref{superpotspp}) we can
choose
\begin{eqnarray}
 x = X_{11} = q_{23}q_{32}  \hbox{ , }
 y = q_{12} q_{21}=q_{13}q_{31}\hbox{ , }
 w = q_{13} q_{32} q_{21} \hbox{ , }
 z = q_{12} q_{23} q_{31} 
\end{eqnarray}
as generators of the mesonic chiral ring. The set of algebraic
relations among these fields reproduces the geometric singularity
(\ref{spp}).  The presence of an adjoint chiral field is a signal
for the presence of a non isolated singularity. In fact, giving a
vev to $X_{11}$ corresponds to motion in the geometry along the $x$
direction, which is a line of non isolated
$\mathbb{C}^2/\mathbb{Z}_2$ singularities: $y^2 = w z$. This line of
singularities can be deformed to a smooth space 
\be
\label{defmetalbe} xy^2=wz \rightarrow xy(y-\xi)=wz  \, .
\ee
%
Moreover
the conical singularity (\ref{spp})
has a complex deformation in which the
tip of the cone is substituted by a three sphere
$S^3$. In this case the
SPP geometry is deformed as
\be
\label{defsusalbe}
xy^2-y\epsilon -wz =0 \, .
\ee
This is the same process as the conifold transition in the
KS solution \cite{Klebanov:2000hb}.

Using toric geometry it is possible to visualize these two
processes. First of all draw the toric diagram of the singularity.
\begin{figure}[h!!!]
\begin{center}
\includegraphics[scale=0.55]{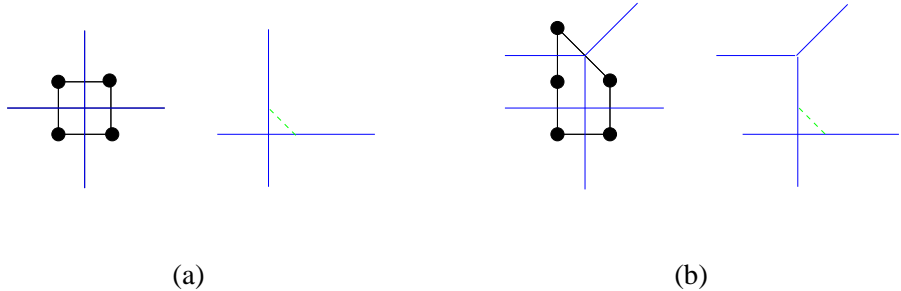}
\caption{Toric diagram, dual diagram and complex deformation for (a)
the conifold case $xy-wz=0 \rightarrow xy-wz -\epsilon =0$; (b) the
SPP case $x y^2- wz=0 \rightarrow xy^2-y\epsilon -wz =0$. The broken
line represents the $S^3$ due to the fluxes. The volume of $S^3$ is
parameterized by $\epsilon$.} \label{defspp}
\end{center}
\end{figure}
Then, if the dual graph has some parallel lines, this implies that
there exist non isolated $\mathbb{C}^2/\mathbb{Z}_k$ lines of
singularities (depending on the number of parallel lines). These
singularities can be deformed by inserting two spheres $S^2$
parameterized by a set of complex $\xi_i$ parameters. If the dual
diagram admits splits in equilibrium (the edges of every
sub-diagrams sum to zero), there exist deformations of the
singularities on the tip of the cone. These deformations are
obtained by inserting three spheres $S^3$, parameterized by some set
of complex $\epsilon_j$ parameters (see Figure \ref{defspp}).

In this paper we argue that metastable supersymmetry breaking is
geometrically realized by moving in the space of complex
deformations. The motion in the $\xi$-parameter space breaks
supersymmetry (in a metastable vacuum) while moving in the
$\epsilon$-parameter space restores the supersymmetry. We will
provide several examples and show that this is a general phenomenon
in an infinite class of
quiver gauge theories.\\

We now review the possible IR behavior of the SPP gauge theory and
their geometric interpretation. The SPP gauge theory has two kinds
of fractional branes, because of the non anomalous distribution of
ranks for the gauge group factors: $(1,0,0)$ and $(0,1,0)$. The
different combinations of these set of branes and the possible
geometric deformations of the singularity characterize different IR
dynamics. We summarize the different possibilities.

The first fractional brane is called an $\mathcal{N}=2$ brane. The
quiver in Figure \ref{tqspp} with $(N_1,0,0)$ fractional branes
reduces to an $\mathcal{N}=2$ gauge theory. The vev of the adjoint
field $X_{11}$ is a modulus of the theory, corresponding to $x$ in
the geometry.
Moving along $x$ corresponds to the D-brane
exploring the curve of $A_1$ singularities $y^2=w z $.

The second fractional brane is called deformation brane. Indeed the
back reaction of $(0,N_2,0)$ $D5$ branes wrapped on the collapsed
two cycle of the conifold inside the $SPP$ induces a geometric
transition which deforms the singularity to a smooth manifold: $x
y^2 + \epsilon y = wz $ (see Figure \ref{defspp}). In the gauge
theory description, the deformation parameter $\epsilon$ is related
to the gaugino condensate. The branes $(0,N_2,0)$ induce deformation
in the geometry and confinement in the gauge theory
\cite{Franco:2005fd}.

The deformation brane and the $\mathcal{N}=2$ brane are
incompatible. If we put $(N_1,N_2,0)$ branes in the $SPP$
singularity the gauge theory has a runaway behavior, which is the
most common behavior in non conformal quiver gauge theories
\cite{Franco:2005zu,Intriligator:2005aw,Brini:2006ej}. Consider the
case $N_2\gg N_1=1 $: the perturbative superpotential is
\begin{equation}
W_{pert}= X_{11} q_{12} q_{21} \, .
\end{equation}
The node 2 is UV free and develops strong dynamics in the IR.
The gauge invariant operators are the
 degrees of freedom that describe the IR dynamics of this
node, i.e. the meson $M_{11}=q_{12}q_{21}$.
The node 2 has
$N_c>N_f $ and generates a non perturbative ADS superpotential.
The complete
IR superpotential is then
\begin{equation}
W_{IR}=X_{11} M_{11}+ (N_2-1) \Big( \frac{\Lambda^{3N_2-1}}{M_{11}}
\Big)^{\frac{1}{N_2-1}} \, .
\end{equation}
The F term equations give the runaway.

Now we can include in the theory the deformation parameter $\xi$ of
the $A_1$ singularity and obtain the geometry (\ref{defmetalbe}).
This corresponds to the superpotential term:
$W_{\xi}=-\xi(X_{11}-q_{13}q_{31})$. Taking the same brane
distribution as in the previous case,
 the IR
superpotential is
\begin{equation}
W_{IR}=X_{11} M_{11}+ (N_2-1) \Big( \frac{\Lambda^{3N_2-1}}{M_{11}}\Big)
^{\frac{1}{N_2-1}} - \xi X_{11}
\end{equation}
and hence the theory develops a supersymmetric vacuum.

\begin{figure}[h!!!]
\begin{center}
\includegraphics[scale=0.55]{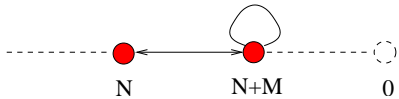}
\caption{The fractional brane disposition to obtain the ISS theory
from the SPP singularity.}
\label{SPPISS}
\end{center}
\end{figure}

Finally, as pointed out in \cite{Buican:2007is}, if we consider the
theory deformed by $\xi$  (\ref{defmetalbe}) in the regime $N_1=N+
M$ and $N_2= N$  (Figure \ref{SPPISS}), the theory admits ISS like
metastable vacua, provided $M>2N$.
%
%
In this case the
 node $N_2$
is the IR free gauge group and the node $N_1$ is treated as the
flavour group (see the Appendix \ref{gaugeunimportant} for a
discussion about this approximation). The superpotential is
\begin{equation}
W_{pert} = - \xi X_{11} + X_{11} q_{12} q_{21}
\end{equation}
and supersymmetry is broken at tree level by the rank condition.
Observe that from this construction we obtain directly the dual
magnetic theory of the ISS model. This theory has also $M$
supersymmetric vacua far away in the moduli space. As usual, these
vacua are obtained by considering the non perturbative contribution
to the superpotential due to the gaugino condensation
\be\label{IRsup} W_{IR} =- \xi X_{11} +N \left(\frac{\det
X_{11}}{\Lambda_m^{2M - N}} \right)^{1/N} \quad \rightarrow \quad
\langle X_{11}\rangle = \Lambda_m^{\frac{M-2N}{M}}\xi^{\frac{N}{M}}
e^{\frac{2\pi \mathrm i k}{M}} \mathbf 1_{M+N} \ee The gauge theory
dynamics that restore supersymmetry have a dual geometric
interpretation. The geometry describing the IR gauge theory is the
$A_1$ deformed conifold variety (\ref{defsusalbe}). Indeed, using
the techniques of \cite{Buican:2007is,Aganagic:2007py}, we can
recover the complete IR non perturbative superpotential
(\ref{IRsup}) from the geometry (\ref{defsusalbe}),
performing a classical computation (see the Appendix \ref{secagana}).\\

The SPP singularity can be considered the simplest representative of
the family of deformable non isolated toric singularities. We will
give a detailed analysis of an infinite sub-class of this family of
singularities called the $L^{aba}$ singularities
\cite{Cvetic:2005ft,Martelli:2005wy,Benvenuti:2005ja,Butti:2005sw,Franco:2005sm},and
we will then comment about their generalizations to more complicated
examples.

\section{The $L^{aba}$ Singularities}\label{labasing}
\begin{figure}[h!!!]
\begin{center}
\includegraphics[scale=0.55]{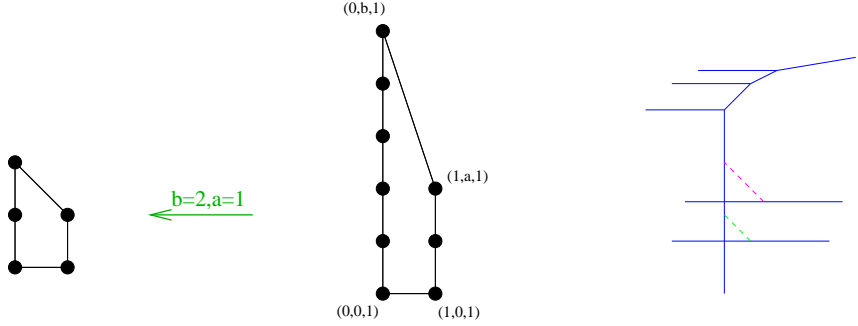}
\caption{Toric diagram of $L^{aba}$ singularity in the case $a=2$,
$b=5$, its dual diagram with the complex deformations, and its
reduction to the $SPP$ toric diagram.}
\label{Lpqq}
\end{center}
\end{figure}

$L^{aba}$ with $b \ge a$ refers to an infinite class of deformable
non isolated singularities that include the $SPP$ as a special case:
$L^{121}=SPP$ (see Figure \ref{Lpqq}).
The $L^{aba}$ singularities contain "$a$" conifold like
singularities (hence "$a$" conifold like complex deformations) and
two lines of non isolated singularities passing through the tip of
the cone: $\mathbb{C}^2/\mathbb{Z}_a$ and
$\mathbb{C}^2/\mathbb{Z}_b$. Indeed the $L^{aba}$ singularities are
described by a quadric in $\mathbb{C}^4$
\begin{equation}\label{Labaeq}
x^a y^b = w z \, .
\end{equation}
The lines parametrized by non zero values of $x$ and $y$ are the
$\mathbb{C}^2/\mathbb{Z}_b$ and $\mathbb{C}^2/\mathbb{Z}_a$ non
isolated singularities
\begin{equation}
\label{nnlocsing}
x\neq 0 \hbox{ } \rightarrow \hbox{ } y^b = w z \hbox{ } \hbox{ , }
\hbox{ } y \neq 0 \hbox{ } \rightarrow \hbox{ } x^a = w z \, .
\end{equation}
We can deform the singularities (\ref{nnlocsing}) by inserting two
cycles at the singular point. A generic $\mathbb{C}^2/\mathbb{Z}_n$
contains, indeed, $n-1$ two spheres collapsed at the origin and can
be deformed to a smooth space turning on $n-1$ generic complex
deformation parameters $\xi_i$, $i=1,...,n-1$
\begin{equation}
x^n = y z \rightarrow x \prod_{i=1}^{n-1} (x-\xi_i)=y z \, .
\end{equation}
On the other hand, from figure \ref{Lpqq}, we note that
 $L^{aba}$ contain $a$ conifolds that can be locally
deformed as
\begin{equation}
x y - w z = 0 \hbox{ } \rightarrow \hbox{ } x y - w z - \epsilon_j = 0
\hbox{ , } \hbox{ } \hbox{ j=1,...a}  \, .
\end{equation}
We have thus identified two families of deformations: the $\xi$
deformations and the $\epsilon$ deformations. As already mentioned,
we argue that the motion in the $\xi$ deformations breaks
supersymmetry to a
 metastable vacuum, while the motion in the $\epsilon$
deformations restores it.

\begin{figure}[h!!!]
\begin{center}
\includegraphics[scale=0.55]{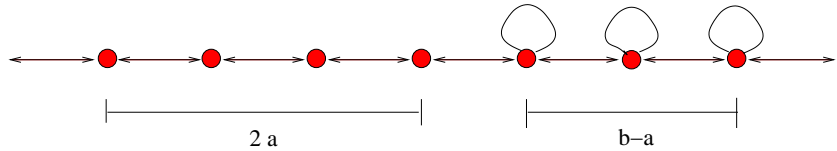}
\caption{The quiver for the generic $L^{aba}$ singularity.}
\label{quivLaba}
\end{center}
\end{figure}

The gauge theories dual to these singularities
\cite{Benvenuti:2005ja,Butti:2005sw,Franco:2005sm} are non chiral
and have the quiver representations\footnote{ This is just a
possible toric phase. By Seiberg duality one can move to other toric
phases with generically different content of matter and different
superpotential but all flowing to the same IR fixed point and hence
having the same singularity as mesonic moduli space.} in Figure
\ref{quivLaba}.
The theory has gauge group $U(N_1) \times U(N_2) \times ... \times
U(N_{a+b})$ and chiral fields transforming in the adjoint or in the
bi-fundamental representations. The superpotentials are
\begin{eqnarray}
\label{Labasup}
W=&&\sum_{i=2a+1}^{b+a} X_{ii} (q_{i,i-1} q_{i-1,i}-q_{i,i+1}
q_{i+1,i}) + \sum_{j=1}^{2a}(-1)^{j+1}q_{j,j-1} q_{j-1,j} q_{j,j+1}
q_{j+1,j}
\end{eqnarray}
where $a+b+1=1$ and the fields $X_{ii}$
transform in the adjoint representation of the i-th gauge group, while
$q_{i,i+1}$ transform in the fundamental representation of the i-th
group and in the anti-fundamental of the i+1-th group.

The chiral ring constrains of the gauge theory can be related to the
algebraic geometric description of the singularity. The complex
deformations can be mapped into deformations of the superpotential,
as well. Indeed the equation (\ref{Labaeq}) can be reconstructed
through the supersymmetric constraints on the mesonic chiral ring of
the gauge theory. Define the following set of basic mesonic chiral
operators
\begin{eqnarray}
& & x_1=q_{12} q_{21} \hbox{ }\hbox{ , }\hbox{ } x_2=q_{34} q_{43}
\hbox{ }\hbox{ , }\hbox{ }\hbox{...} \hbox{ }\hbox{ , }\hbox{
}x_a=q_{2a-1,2a} q_{2a,2a-1} ;\nonumber\\ & & y_1=q_{23} q_{32} \hbox{
}\hbox{ , }\hbox{ } y_2=q_{45} q_{54} \hbox{ }\hbox{ , }\hbox{
}\hbox{...} \hbox{ }\hbox{ , }\hbox{ }y_a=q_{2a,2a+1} q_{2a+1,2a}, \\
\nonumber & & y_{a+1}=q_{2a+1,2a+2} q_{2a+2,2a+1} \hbox{ }\hbox{ ,
}\hbox{ }\hbox{...} \hbox{ }\hbox{ , }\hbox{ } y_{b}=q_{b,1}
q_{1,b};\nonumber\\ & & X_{2a+1,2a+1}\hbox{ }\hbox{ , }\hbox{
}X_{2a+2,2a+2}\hbox{ }\hbox{ , }\hbox{ }\hbox{...} \hbox{ }\hbox{ ,
}\hbox{ } X_{b,b}; \nonumber\\ & & w= q_{1,b} q_{b,b-1}\hbox{
}\hbox{...} \hbox{ }q_{3,2}q_{2,1}\hbox{ }\hbox{ , }\hbox{ }z= q_{1,2}
q_{2,3}\hbox{ }\hbox{...} \hbox{ }q_{b-1,b}q_{b,1} \, .
\end{eqnarray}
These operators satisfy
\begin{equation}
\label{eq1}
x_1...x_a \hbox{   }  y_1...y_b = w z \, .
\end{equation}
From the $F$-term equations we get the relations
\begin{equation}\label{eq2}
x_1 = ... = x_a = X_{2a+1,2a+1} = ... = X_{b,b} \hbox{ }\hbox{ ,
}\hbox{ } y_1= ... = y_b \, .
\end{equation}
The chiral ring constraints (\ref{eq1},\ref{eq2}) reproduce the
geometric singularity (\ref{Labaeq}).

By this technique, using the $F$-term constraints,
 we can also map the
complex deformations of the geometry to deformations of the
superpotential.

A final, important, remark is that different UV gauge theories,
flowing in the IR to the same conformal fixed point, correspond to
the same toric singularity. These theories are related by Seiberg
dualities and give equivalent physical descriptions at the conformal
point. In this paper we choose the more convenient Seiberg phase for
finding metastable vacua in the related non conformal case. This can
be achieved by performing a set of Seiberg dualities on the quiver
gauge theories with only regular branes, and then placing the right
set of fractional branes that breaks conformal invariance.

\section{Meta-stable vacua in $L^{aba}$ theories}\label{metlaba}

This section is devoted to the analysis of metastability in the
$L^{aba}$ theories with $b>a$. The simplest example is the one
studied in \cite{Buican:2007is}, where the ISS dynamics dynamics was
found in the infrared of a deformed $L^{121}$ theory. We now extend
this analysis to more complicated cases, like $L^{131}$ and then
$L^{1n1}$. After that we show how to generate chains of theories
that have supersymmetry breaking meta-stable vacua. Generally
speaking, if we have a $L^{aba}$ theory which shows metastability,
we argue that the $L^{an,bn,an}$ theory behaves as a set of
decoupled theories of this sort. At the end of this section, we give
a general recipe for the existence of metastable vacua in a
$L^{aba}$ theory, by decomposing it into a set of shorter quivers.

In the analysis of the metastable vacua we consider some nodes of
the quivers as gauge groups and other nodes as flavor groups, tuning
the dynamical scales as explained in the Appendix
(\ref{gaugeunimportant}). This is implicit in all the cases that we
treat.

Note that, in the notation of ISS, we are working in the magnetic
description. This means that we deal with IR free gauge groups,
without performing Seiberg duality on them. Another important remark
is that, since we are dealing with the magnetic phase, if we want to
realize metastable vacua, we need linear deformations in the mesons
rather than massive quarks.

We present here several examples, as well as general results,
 to stress the fact that the $\xi$ deformations lead to metastable non
supersymmetric vacua whereas the $\epsilon$ deformations bring to
supersymmetry restoration. We leave the details of the field theory
analysis in the Appendix \ref{appB}.

\subsection{The $L^{131}$ theory} \label{Lunotreuno}
\begin{figure}[htpb]
\begin{center}
\includegraphics[width=5cm]{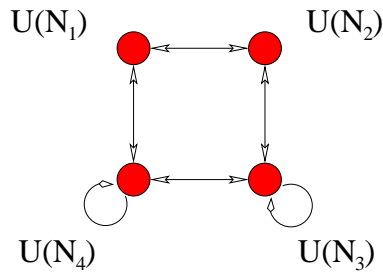}
\caption{Quiver for the $L^{131}$ theory.}
\label{l131}
\end{center}
\end{figure}

The $L^{131}$ theory is described by the quiver in
figure \ref{l131},
with superpotential
\bea
\label{dfs}
W=&& X_{33} (q_{32} q_{23}-q_{34} q_{43})
-h q_{21} q_{12} q_{23} q_{32}+ h  q_{12}q_{21} q_{14}q_{41} +\\
&&+X_{44} (q_{43} q_{34}-q_{41}q_{14} )
\eea
and it corresponds to the singular geometry
\be
x y^3 = w z
\ee
which is correctly reproduced by the mesonic chiral ring
as explained in section 2.

We now add a superpotential deformation
\be
\label{dsf}
W_{def}= - \xi_1 ( X_{33} - h q_{12} q_{21})
 -\xi_2 (X_{44} - h q_{12} q_{21}) \, .
\ee
 Imposing the constraints from the $F$-term equations we find the
new relations on the mesonic chiral ring 
\be y = q_{23}q_{32} =
q_{34}q_{43}+\xi_1 = q_{41}q_{14}+\xi_1+\xi_2 \, . 
\ee 
These constraints
are translated into the deformed geometry 
\be x y
(y-\xi_1)(y-\xi_1-\xi_2)  =wz \, .
\ee 
Obviously, we are not obliged to
add a linear term for each adjoint field but the case with only one
deformation turns out to be unstable, as we show in the following.

We study this theory setting one node to zero. There are two
different possibilities: we can set to zero a node with an adjoint
field ($N_3$ or $N_4$) or a node without it ($N_1$ or $N_2$),
obtaining a theory with one or two adjoint fields respectively. In
the second case the scalar potential has dangerous flat directions
and we cannot find metastable vacua. In the following we only
analyze the first case and show the existence of long living non
supersymmetric metastable vacua.

\begin{figure}[h!!!]
\begin{center}
  \includegraphics[width=4cm]{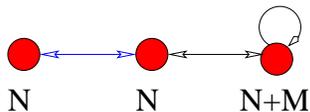}
\caption{$L^{131}$ theory with $N_4=0$. The blue lines indicate
massive matter.}
\label{l131a}
\end{center}
\end{figure}

The theory under investigation is then obtained
setting to zero the $N_4$ node (the case with $N_3=0$ is the same),
described by the
quiver in figure \ref{l131a}.
\\
The superpotential is 
\be W= X_{33} q_{32} q_{23}- h q_{21} q_{12}
q_{23} q_{32} -\xi_1 X_{33} +h (\xi_1 +\xi_2)   q_{21} q_{12} \, . 
\ee
For simplicity, in the analysis of the equations of motion we fix
the ranks of the groups to be 
\be 
N_1 = N_2 =N \ \ \ \ \ N_3 = N+M \, .
\ee

First of all we have to impose the correct tuning on the scales of
the gauge groups and on the rank numbers in order to treat the node
$N_2$ as an infrared free gauge group and the other gauge groups as
flavours. Calculating the beta functions we have 
\be b_1 = 2N \ \ \
\ b_2=N-M \ \ \ \ b_3=N+2M \, .
\ee 
Since we require the group $U(N_2)$
to be infrared free we impose the constraint $M>N$. Moreover, we
require that this group is more coupled than the other groups at the
supersymmetry breaking scale and at the scale of supersymmetry
restoration\footnote {With supersymmetry restoration we mean the
supersymmetric vacua that arise due to the strong dynamics of
$U(N_2)$. For what concern the other supersymmetric vacua, given by
the strong dynamics of the other groups, the tuning on the scales
put them far away in the field space.}. This can be done by tuning
the scales $\Lambda_1$ and $\Lambda_3$, which are the strong
coupling scales of two UV free gauge groups. Their scales have to be
chosen\footnote{ See Appendix \ref{gaugeunimportant} and
\cite{Amariti:2007am} for a complete analysis. } much smaller than
the scale of supersymmetry breaking (which is the deformation $h
\xi_1$) and much smaller than the scale $\Lambda_2$ of $U(N_2)$.

Now that we have correctly set up
the role played by each gauge group in the quiver
we can proceed in finding the vacua.
A detailed analysis of this theory is left to the Appendix \ref{luciano1}.
Here we sketch the main results.
The $F$-term equation for the $X_{33}$ field is the rank condition
and breaks supersymmetry,
fixing the vev of the fields
$q_{23}$ and $q_{32}$.
The equation for the $q_{12}$
quark is
\be
\label{eqquark131}
F_{q_{12}}=h \left( - q_{23}q_{32}+(\xi_1+\xi_2)\right)q_{21}=
h \xi_2 q_{21}
\ee
and is solved with $q_{12}=0=q_{21}$.
This is related to the fact that we have
added two deformation parameters ($\xi_1$, $\xi_2$), i.e. two
linear contributions to the superpotential
for 
the two adjoint fields.
Otherwise (for $\xi_2=0$),  the equation (\ref{eqquark131})
would be automatically satisfied,
leaving the fields $q_{12}$ and $q_{21}$
unfixed at tree level and leading to
potentially dangerous flat directions.
%
%

%
%
The non supersymmetric vacuum at tree level
is
\be
q_{12}=q_{21}^{T}=0 \qquad
q_{32}=q_{23}^{T}=\left(
\begin{array}{c}
\sqrt{\xi_1}
\mathbf{1}_{N} \\
0
\end{array}
\right)
\qquad
X_{33}=
\left(
\begin{array}{c c}
0 & 0\\
0 &\chi
\end{array}
\right)
\ee
where $\chi$ is the pseudomodulus of dimension $M \times M$.
As outlined in the Appendix \ref{luciano1}
this vacuum is stable under one loop correction,
and the pseudomodulus is stabilized at $\chi=0$.

The restoration of supersymmetry is obtained 
in the hypothesis that the group labeled by $N_2$ develops a strong
dynamics, by adding to the low energy superpotential a non
perturbative contribution
\be
\label{fieldthspot}
W_{IR} = - \xi_1  X_{33} + N_2 \left(\Lambda^{3N_2 -
N_3}\det{X_{33}}\right)^{\frac{1}{N_2}}
\ee
where we have integrated out all the massive fields.
From the geometric point of view, supersymmetry restoration,
governed by the dynamics of the $U(N_2)$ gauge group, can be
described deforming the geometry with an $S^3$, i.e. an $\epsilon$
deformation, 
\be\label{susyl131} 
(y-\xi_1) (y-\xi_1-\xi_2)
(xy-\epsilon)=wz  \, .
\ee
The low energy field theory superpotential (\ref{fieldthspot}) can
be recovered from the geometric data (\ref{susyl131}). Indeed,
setting
$y=x'-y'$ and $x=x'+y'$, 
equation (\ref{susyl131}) becomes: 
\be
(x'-y'-\xi_1)(x'-y'-\xi_1-\xi_2)
\left(x'+\sqrt{y'^2+\epsilon}\right)
\left(x'-\sqrt{y'^2+\epsilon}\right) =wz \, .
\ee 
The low energy
superpotential can be written as a function of the glueball field
$S_2$ (identified with $\epsilon/2$)
and of the adjoint field $X_{33}$ 
\be
W_{IR}(S_2,X_{33}) = W_{GVW}(S_2)+W_{adj}(S_2,X_{33})= N_2 S_2
\left(\log\frac{S_2}{\Lambda_2^3}-1\right) +\frac{t}{g_2} S_2 +
W_{adj}(S_2,X_{33}) \, .
\ee
Following the procedure explained in
Appendix \ref{secagana} the last term is derived from the
geometric data
\bea
W_{adj}(S_2,X_{33})&& = \int
({x'}_2(y')-{x'}_3(y'))dy'=
\\
&&=\int \left(
y'+{\xi_1}-\sqrt{y'^2+\epsilon}
\right)
\sim
\xi_1 X_{33}-
S_2 \log\frac{X_{33}}{\Lambda_m}
\nonumber
\eea
where
we have expanded the integral in the approximation
$y' \gg \xi_1, \epsilon$.
We can now solve the
equation of motion
for the glueball
field $S_2$ and integrate it out, ignoring the multi-istanton contribution.
In this way
we recover
from the geometry (\ref{susyl131})
the low
energy superpotential (\ref{fieldthspot}).

As claimed in the introduction,
we
have shown, in this simple example,
 that the $\xi_i$ deformations
lead to metastable vacua
whereas the $\epsilon$ deformation
leads to supersymmetry restoration.

\subsection{The $L^{1n1}$ theories}
The metastable $L^{131}$ theory can be generalized to the more
complicate $L^{1n1}$ case. We find metastable supersymmetry breaking
in the $L^{141}$ and $L^{151}$ theories and then we show how to
extend this procedure to the $L^{1n1}$ case. A relevant aspect in
the analysis is the decoupling between the breaking sector and the
supersymmetric one. Once again we leave the details in the Appendix
\ref{appB}, being here mainly interested in the geometrical
realization of metastable supersymmetry breaking.

\subsubsection{$L^{141}$}
\label{secl141} 
\begin{figure}[h!!!]
\begin{center}
  \includegraphics[width=6cm]{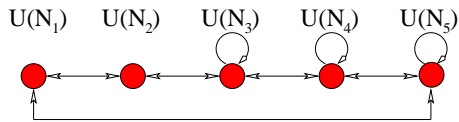}
\caption{$L^{141}$ theory.}
\label{L141}
\end{center}
\end{figure}

Here we study the quiver gauge theory of figure
\ref{L141},
with superpotential
\bea W= &&h q_{12} q_{23} q_{32} q_{21} -X_{33} q_{32} q_{23}+X_{33}
q_{34} q_{43}-X_{44} q_{43} q_{34}\nonumber \\
&& X_{44} q_{45}
q_{54}-X_{55} q_{54} q_{45} + X_{55} q_{51} q_{15}- h q_{51} q_{12}
q_{21} q_{15} \, .
\eea
The geometry associated with this theory is described by the
equation 
\be 
x  y^4  = w z \, .
\ee 
Supersymmetry breaking is driven by
linear terms for the adjoint fields. We add the deformation
superpotential
\be W_{\text{def}} = - \xi_3( X_{33}-h q_{32}q_{23})+
\xi_4(X_{44}-h q_{32}q_{23})\ -\xi_5(X_{55}-h q_{32}q_{23}) \, .
\ee
We have add a linear term for all the adjoint fields: this is
crucial for the stability of the non supersymmetric vacuum. The
$q_{23}$ and $q_{32}$ quarks become massive, since the $F$-terms
constraints have to be compatible.

The corresponding geometry reads now
\be 
x (y-\xi_3)
(y-\xi_3-\xi_4)(y-\xi_3 -\xi_4-\xi_5) y =wz \, .
\ee
If we consider as gauge group the node $U(N_2)$ and choosing the
ranks as\footnote{Note that also the situation with gauge group
$U(N_1)$ and $N_3=N$ and $N_5=N+M$ leads to metastable vacua.} 
\be
N_1=N_2=N_5=N \quad N_3=N+M \quad N_4=0 
\ee 
with $M>N$, this theory
breaks supersymmetry through rank condition for the meson $X_{33}$.
A detailed analysis (see Appendix \ref{luciano2}) shows that this
theory possesses metastable vacua without dangerous flat directions.

Two important remarks are in order.
Without turning on the
deformation $\xi_4$ (the one related to the node set to zero) we are
not protected from instabilities of
the scalar potential (see Appendix \ref{luciano2}).
Furthermore, as we did in the $L^{131}$ case, we have decoupled an
ISS like sector with supersymmetry breaking from a supersymmetric
sector.
These two facts hold in all the $L^{1n1}$ cases.
%

The process of supersymmetry restoration works as
in the $L^{131}$,
when the dynamics of the
gauge
group $U(N_2)$
gives rise to non perturbative contributions to the superpotential.

\subsubsection{$L^{151}$}
\label{secl151}
Here we study
metastability in the $L^{151}$ quiver gauge theory. This is the
basic example for the generalization of the analysis to the
$L^{1n1}$ case.
\begin{figure}[h!!!]
\begin{center}
\includegraphics[width=6cm]{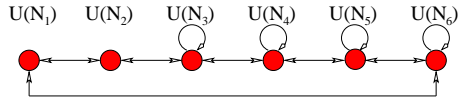}
\caption{The $L^{151}$ theory.}
\label{L151}
\end{center}
\end{figure}
The gauge theory, related to the quiver in figure
\ref{L151},
has superpotential
\bea
W= && h q_{12} q_{23} q_{32} q_{21} -X_{33} q_{32} q_{23}+X_{33}
q_{34} q_{43}-X_{44} q_{43} q_{34} +X_{44} q_{45}
q_{54}\nonumber \\
&& -X_{55} q_{54} q_{45} + X_{55} q_{51} q_{15}
-X_{66} q_{65} q_{56} + X_{66} q_{61} q_{16}
- h q_{61} q_{12} q_{21} q_{16}
\eea
and it is associated to the geometry
\be
x y^5  = w z \, .
\ee
Once again we deform the superpotential
with linear terms for the adjoint fields and masses
for the quarks
\be
\label{defL151}
W_{\text{def}} = -\xi_3(X_{33}-h q_{32}q_{23})
-\xi_4(X_{44}-h q_{32}q_{23})
-\xi_5(X_{55}-h q_{32}q_{23})
-\xi_6(X_{66}-h q_{32}q_{23}) \, .
\ee
The deformation (\ref{defL151}) leads to the geometric deformation
\be 
x (y-\xi_3) (y-\xi_3-\xi_4)(y-\xi_3
-\xi_4-\xi_5)(y-\xi_3-\xi_4-\xi_5 -\xi_6) y =wz \, .
\ee
We choose the ranks of the groups as 
\be 
N_1 = N_2 =N_5 =N_6 = N
\qquad N_3 = N+M \qquad N_4=0 
\ee
with $M>N$. 
The equation of motion for the field
$X_{33}$ is the ISS rank condition, that breaks supersymmetry at the
classical level.
%
In the Appendix \ref{luciano3}
we show that the supersymmetry breaking minimum is stable.
Stability of the metastable vacuum
requires $\xi_3,\xi_4,\xi_5 \neq 0$
and arbitrary $\xi_6$.

The supersymmetry restoration carries on exactly as
in the $L^{131}$,
with non perturbative contribution
to the superpotential
due to the dynamics of the gauge group $U(N_2)$.


\subsubsection{$L^{1n1}$}
We now extend the results about metastability to the
general $L^{1n1}$ theory.
The
superpotential is
\bea
W=&&\sum_{i=3}^{n} X_{i,i} (q_{i,i-1} q_{i-1,i}-q_{i,i+1} q_{i+1,i})
+h q_{21} q_{12} q_{23} q_{32}- h q_{12}q_{21} q_{1,n+1}q_{n+1,1} +\\
&&+X_{n+1,n+1} (q_{1,n+1} q_{n+1,1}-q_{n,n+1}q_{n+1,n} )
\eea
and the geometry
\be
x y^n =wz \, .
\ee
The deformation of the superpotential is
\be
\label{supdef}
\Delta W=\sum_{i=3}^{n+1}\xi_i ( h q_{12} q_{21} - X_{i,i} )
\ee
which corresponds to the geometry
\be
x y \prod_{i=3}^{n+1}(y-\sum_{j=1}^{i}\xi_i) =wz \, .
\ee
We choose the
 ranks of the nodes to be
 \be
 N_4=0 \qquad
 N_3=N+M \qquad
 N_j=N \quad (j \neq 3,4)
 \ee
 such that supersymmetry is broken at node 3.
 Moreover it should be $M>N$ for $U(N_2)$
 to be IR free.

 The deformations $\xi_3$, $\xi_4$ and $\xi_5$ have to be different from zero
 for the non supersymmetric vacuum to be stable.
All the other deformations can be chosen arbitrarily (see Appendix
\ref{luciano4}).
The breaking sector is the same than all the other $L^{1n1}$
cases
analyzed before. The only difference is that the supersymmetric
sector is larger here.

Supersymmetry is restored by the strong dynamics of the gauge group
$U(N_2)$, and the metastable vacuum is long living.
This concludes the analysis of the $L^{1n1}$ theories.

\subsection{Extension to longer quivers}
We extend here the analysis of the
$L^{1n1}$ theories to more complicated $L^{aba}$ cases.
The strategy is to decouple an $L^{aba}$ theory in a set of $a$
metastable theories, adding $b-a$ deformations, one for each adjoint
field.  In the case $b-a \geq a$ ,by using the results obtained for
$L^{1n1}$, we are able to find metastable vacua in each $L^{aba}$
theory.

Our general strategy will be to consider in each metastable subset
only one group as a gauge group, since there are some difficulties in
treating the dynamics of more than one gauge group simultaneously.

We study first the simplest cases, like the $L^{n2nn}$ theory, which
can be viewed as a set of decoupled ISS models.  This is a
pedagogical example, useful for the extension of the analysis to the
general situations and for the proof that metastability is a generic
phenomenon in the $L^{aba}$ theories. At the end of this subsection,
we furnish the general recipe to build metastable $L^{aba}$ quivers.

\subsubsection{$L^{n2nn}$ as a set of decoupled ISS}
\begin{figure}
\begin{center}
\begin{tabular}{cc}
\includegraphics[width=10cm]{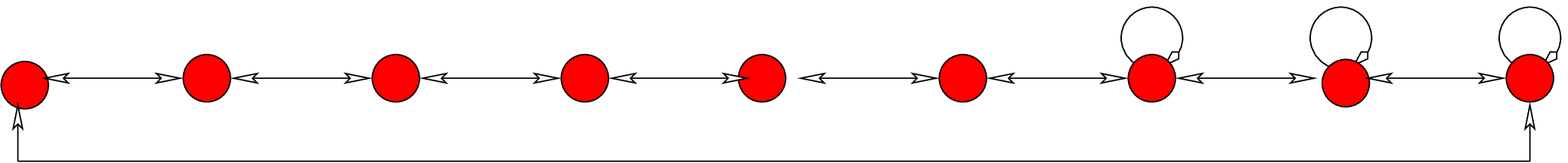}
\\
(a)
\\
\\
\includegraphics[width=10cm]{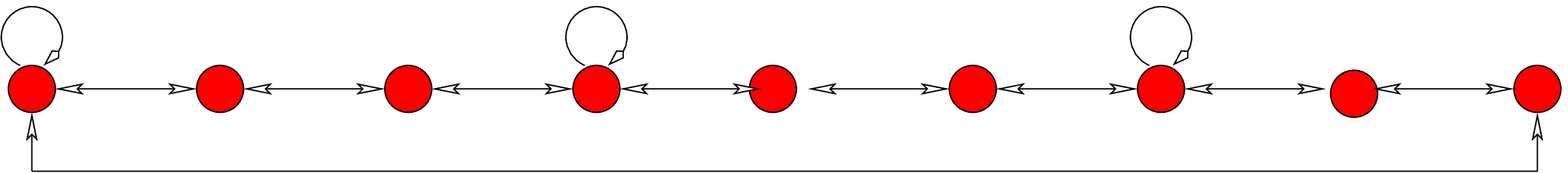}
\\
(b)
\end{tabular}
\end{center}
\caption{Two different Seiberg phases of the same $L^{363}$ quiver
gauge theory. }
\label{L363}
\end{figure}

The $L^{121}$ gauge theory, in the ISS regime, has been shown to possess
meta-stable vacua \cite{Buican:2007is}.
%
%
Starting from an $L^{n2nn}$ quiver gauge theory
we can perform a set of Seiberg dualities going from the
first quiver in the Figure \ref{L363} to the second one.
In fact, Seiberg duality on these theories
has the effect of
displacing the adjoint fields.
Each duality
moves one adjoint field
two nodes farther.

\begin{figure}[h!!!]
\begin{center}
\begin{tabular}{cc}
\includegraphics[width=10cm]{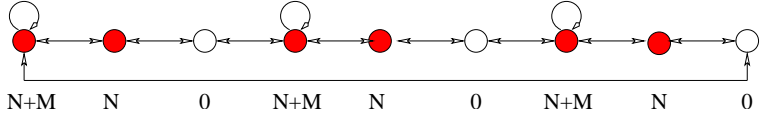}
\end{tabular}
\caption{$L^{363}$ as a set of three decoupled ISS models}
\label{dueissdis}
\end{center}
\end{figure}

We now deform the geometry, associating each $\xi_i$ deformation
with the $i$-th node, obtaining 
\be
 \prod_{i=1}^{n} x (y-\xi_{3i-2})
y  =wz  \, .
\ee
 This deformation corresponds, on the gauge side,
to the combined addition of linear terms for the adjoint fields and
of masses for the appropriate bifundamentals (i.e. that ones not
directly coupled to the adjoint fields). By setting to zero one
node, without an adjoint field, every three nodes, we have a theory
of decoupled metastable ISS models (see Figure \ref{dueissdis}).
The analysis of metastable vacua is the same as in ISS for each
sector. Supersymmetry is restored in the large field region
%
in each ISS sector,
where the gauge group gives rise to a non
perturbative contribution 
in the effective theory.
The non perturbative contributions modify the constraints
on the mesonic moduli space,
and hence the geometry, as
\be
\prod_{i=1}^{n} y \left(  (y-\xi_{3i-2})x-\epsilon_i \right) =wz \, .
\ee

The technique of
Appendix \ref{secagana}
can be applied to the new singularities
of the geometry
to recover the correct low
energy behavior of the field theory.
The calculation proceeds
exactly as in the $L^{121}$ case

\subsubsection{The $L^{n3nn}$ theories}
\begin{figure}
\begin{center}
\begin{tabular}{cc}
\includegraphics[width=12cm]{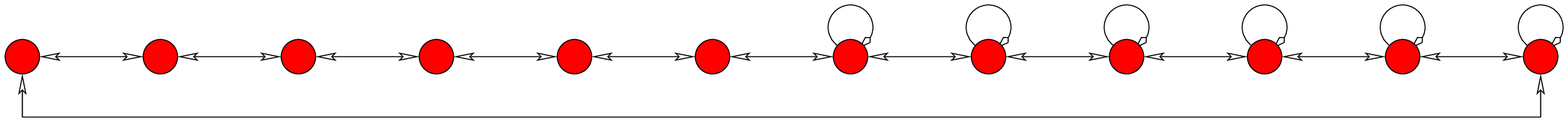}
\\
(a)
\\
\\
\includegraphics[width=12cm]{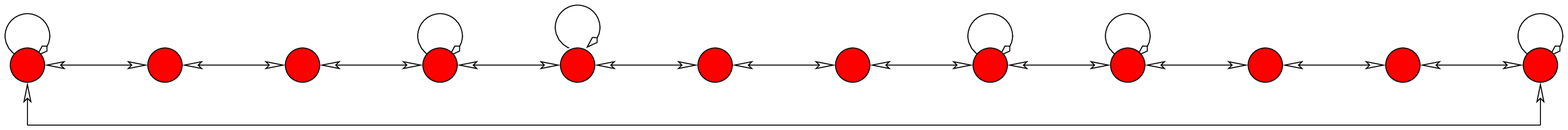}
\\
(b)
\end{tabular}
\end{center}
\caption{Two different Seiberg phases of the same $L^{393}$ quiver
gauge theory. }
\label{Ln3nn}
\end{figure}

With this strategy
we can build longer quivers with metastable vacua.
For example the $L^{131}$ case
can be extended to metastable $L^{n3nn}$
theories.
Indeed we can perform a set of Seiberg dualities to
obtain a new phase of the theory as shown in
Figure \ref{Ln3nn}.
As we did in the $L^{n 2n n}$ case,
we then deform the geometry
\be
\prod_{i=1}^{2n}(y-\xi_{4(i-1)}-\xi_{4(i+1)+1})y x  (y-\xi_{4i})=wz
\ \ \ \ \ \ \ \ \ \ \
\text{with} \ \ \  \xi_0=\xi_{4n} \, .
\ee
%
The deformation brings in
the superpotential a linear term for each adjoint field,
and a mass term for the quarks stretched between two nodes without
the adjoint fields.

\begin{figure}[h!!!]
\begin{center}
\includegraphics[width=15cm]{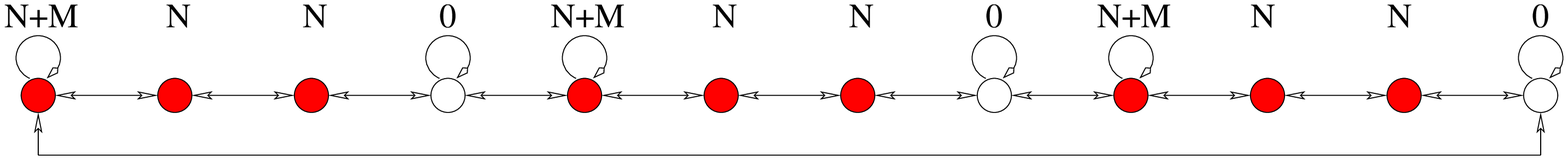}
\caption{$L^{393}$ as a set of three decoupled $L^{131}$ models.}

\label{Ln3nnbis}
\end{center}
\end{figure}

We set then to zero the right nodes and breaks
the $L^{n 3n n}$ into a set of metastable gauge theories.
Indeed,
setting the ranks number as in Figure \ref{Ln3nnbis},
in each decoupled sector we have the same breaking
patterns as in the $L^{131}$ studied before.
Each sector has the superpotential
\be
W = h q_{i,i+1} q_{i+1,i+2} q_{i+2,i+1} q_{i+1,i} \
      - q_{i+1,i} X_{i,i} q_{i,i+1}
      -\xi_{i} X_{i,i}  + h (\xi_{i}+\xi_{i+3}) q_{i+2,i+1}q_{i+1,i+2}
\ee
which leads to
long living metastable vacua, as it has been
explained for the $L^{131}$ theory.

Supersymmetry restoration can be obtained separately in
each decoupled sector, through the strong dynamics of
the gauge group.
In the geometric description it can be
read from the deformation of the variety
\be
\prod_{i=1}^{2n}\left((y-\xi_{4(i-1)}-\xi_{4i-3})x-\epsilon_i\right)
y  (y-\xi_{4i})=wz \, .
\ee
It is straightforward to show that it corresponds
to adding a term proportional to $\det X_{ii}$
for each gauge group and restores supersymmetry.


\subsubsection{Extension with an example}
The procedure just outlined for the
$L^{1 2 1}$ and $L^{1 3 1}$
can be applied also for the $L^{1n1}$ case,
extending it to $L^{m, nm, m}$ metastable theories.

More generally,
we can consider an
$L^{aba}$ quiver that can be decomposed into subsets of different
theories, each one metastable.

\begin{figure}[h!!!]
\begin{center}
\begin{tabular}{c}
\includegraphics[width=8cm]{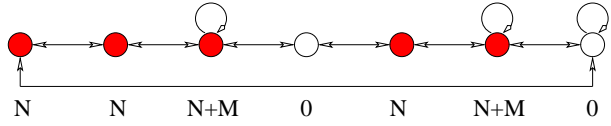}
\end{tabular}
\caption{The Seiberg phase of $L^{252}$ suitable for metastable vacua}
\label{L252fig}
\end{center}
\end{figure}

We show the technique in a clarifying example and then
give a general recipe.
For instance,
we take the $L^{252}$ theory and perform a Seiberg duality
to obtain the phase of
figure \ref{L252fig}.
By deforming all the adjoint fields with a linear term
the chiral ring gets modified to be
\bea
&&y = q_{23}q_{32}=q_{34}q_{43}+\xi_3=
q_{56}q_{65}=q_{67}q_{76}+\xi_6=q_{71}q_{17}+\xi_6+\xi_7
\nonumber \\
&&x = q_{12}q_{21}=q_{45}q_{54} 
\eea 
with the corresponding deformed
geometry 
\be 
x^2 y^2 (y-\xi_3)(y-\xi_6)(y-\xi_6-\xi_7)=wz \, .
\ee 
We
choose then the sequence of the ranks of the groups as shown in
figure \ref{L252fig}, setting to zero the fourth and the last node.
Now the first sector corresponds to the $L^{131}$ theory and the
second one to the $L^{121}$ one. Each sector shows metastable
supersymmetry breaking vacua. The superpotential is 
\be 
W = h q_{12}
q_{23}q_{32} q_{21} - q_{23}X_{33}q_{32} - q_{56}X_{66} q_{65}
-\xi_3 X_{33} -\xi_6 X_{66} + (\xi_3+\xi_7) q_{21}q_{12}  \, .
\ee

Supersymmetry is restored by the strong dynamics
of the nodes
two and five
that give rise to the non perturbative contribution
\be
W_{\text{dyn}}  =
N \left(\Lambda_2^{2N-M}\det X_{33}\right)^{1/N}
+N \left(\Lambda_5^{2N-M} \det X_{66}\right)^{1/N}
\ee
which deforms the geometry to
\be \label{geomeanto}
(x y - \epsilon_1)
(x y - \epsilon_2)
(y-\xi_3)
(y-\xi_6)
(y-\xi_6-\xi_7)
=
wz \, .
\ee
Indeed,
from this geometry, with the technique discussed in the
Appendix \ref{secagana},
we can recover now the low energy superpotential of the field theory.


We start writing the general $IR$ superpotential as a function of
the mesons $X_{33}$ and $X_{66}$ and of the glueballs $S_2$ and
$S_5$ 
\bea 
W_{IR}=W_{GVW}(S_2) + W_{GVW}(S_5) +W_{adj}(S_2,X_{33})
+W_{adj}(S_5,X_{66}) \, .
\eea
 Substituting $y=x'-y'$ and $x=x'+y'$ in
(\ref{geomeanto}) we can calculate the contributions $W(S,X)$ in the
superpotential
\bea 
W_{adj}(S_2,X_{33})&=& \int \left(
y'+\xi_3-\sqrt{{y'}^2+\epsilon_1} \right)dy'\sim \xi_3 X_{33}
-S_2\log{\frac{X_{33}}{\Lambda_2}}
\nonumber \\
W_{adj}(S_5,X_{66})&=&
\int
\left( y'+\xi_6-\sqrt{{y'}^2+\epsilon_2}
\right)dy'\sim \xi_6  X_{66}
-S_5\log{\frac{X_{66}}{\Lambda_5}}
\eea
where we identify $(2S_2,2S_5)=(\epsilon_1,\epsilon_2)$.
We
remark that the variable $y'$, that parametrizes the position of the
brane, can be interpreted as the vev of the field $X_{ii}$ in each
integral.
Integrating out the glueball fields $S_2$ and $S_5$
we recover the low energy description of the field theory.  \\
This example shows that 
we can obtain metastable
$L^{aba}$ theories
by breaking them up into shorter quivers.

\subsubsection{General analysis}

Here we decompose an $L^{aba}$ theory into a set of $L^{1n_i 1}$
theories, each one with metastable vacua.

We consider a distribution of gauge groups with ranks such that
there are no consecutive nodes set to zero.
Moreover we consider only Seiberg phases
with $b-a$ adjoint fields
to be distributed on the $a$ gauge
nodes. This implies that we can only
describe theories with $b-a \geq a$.
In the next section
we extend this result to theories with
$b-a < a$, studying
Seiberg phases with more adjoint fields.

With these assumptions, starting from a $L^{aba}$ and setting $a$
nodes to zero, we can obtain $a$ metastable $L^{1 n_i 1}$ theories.
Each decoupled sector possesses long living metastable vacua like
the ones studied in the $L^{1n_i1}$ theories and hence the whole
theory is metastable.
The procedure is not unique:
we can indeed decouple the $L^{aba}$ theory in different
sets of $L^{1n_i1}$ quivers.
This is related to the fact that we can
distribute differently the $b-a$
adjoint fields on the $a$ gauge nodes
and set to zero nodes with or without adjoint fields.

\begin{figure}[h!!!]
\begin{center}
\includegraphics[width=12cm]{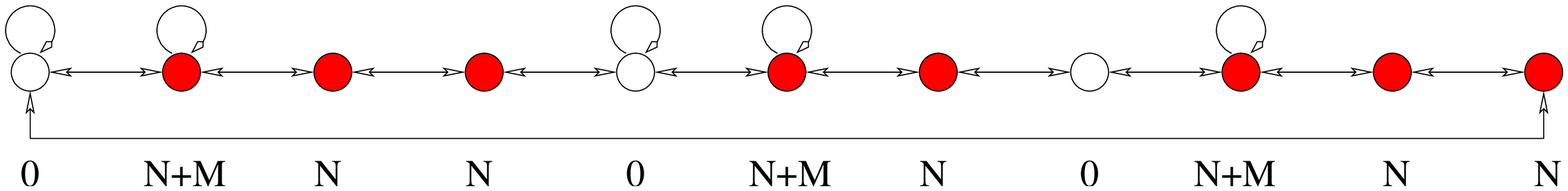}
\\
\includegraphics[width=12cm]{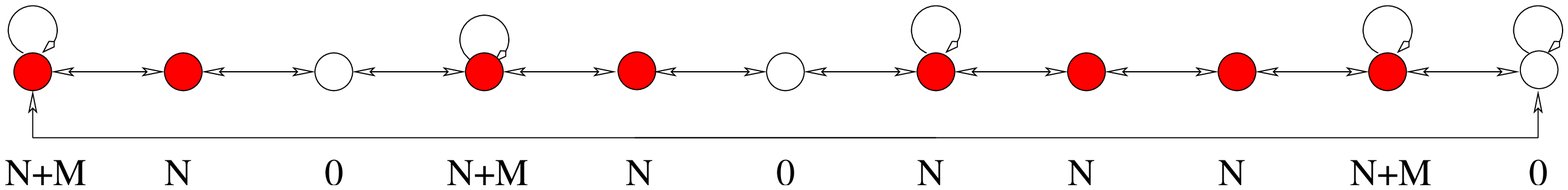}
\caption{The two inequivalent possible $L^{383}$ that give rise to
three decoupled metastable sectors}
\label{antoto}
\end{center}
\end{figure}

This can be shown in a simple example.
The $L^{383}$ theory can be decoupled in
three different sectors, where the number of adjoint fields
totals up to five.  There are two inequivalent possibilities
to obtain metastable vacua
as shown in figure
\ref{antoto}.
We set three different nodes 
to zero (nodes $1,5,8$ in the first case and $3,6,11$ in the second
one), obtaining three decoupled metastable theories. For the first
case the analysis of metastability follows from $L^{121}$ and
$L^{131}$, while in the second case it follows from $L^{121}$ and
$L^{141}$. So we decouple $L^{383}$ in two different ways: as $2
L^{131} + L^{121}$ or as $2L^{121} + L^{141}$.
By this technique,
we can write $L^{aba}$ as a sum of $\sum_{i=1}^a L^{1n_i1}$,
with the constraint $\sum_{i=1}^a n_i = b$, $n_i \geq 2$.
All these theories lead, with the right
distribution of ranks, to metastable vacua.

\section{Meta-stable vacua in $L^{aaa}$ theories}
\label{metlaaa}
In the case $a=b$, i.e. $L^{aaa}$ the theory does not posses adjoint
matter, since $b-a = 0$. Nevertheless, by performing Seiberg
dualities, we can create the necessary adjoint fields.
As explained at the end of section \ref{labasing}, this procedure
does not affect the geometry, which is of the form 
\be 
x^a y^a = w z \, .
\ee 
We can then add the deformations for the adjoint fields and
obtain theories suitable for metastable supersymmetry breaking.

Once again the strategy to analyze a long quiver consists in
breaking it up in a set of shorter quivers, each one with metastable
vacua.

We study in detail the simplest example, $L^{222}$,
and then we comment on possible generalizations.

\subsection{The $L^{222}$ theory}
\begin{figure}[h!!!]
\begin{center}
\includegraphics[width=5cm]{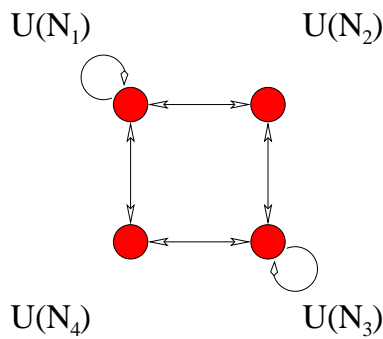}
\caption{The $L^{222}$ quiver without any node set to zero}\label{l222quivero}
\end{center}
\end{figure}

\begin{figure}[h!!!]
\begin{center}
\begin{tabular}{ccc}
~~~~~~~~~
\includegraphics[width=3cm]{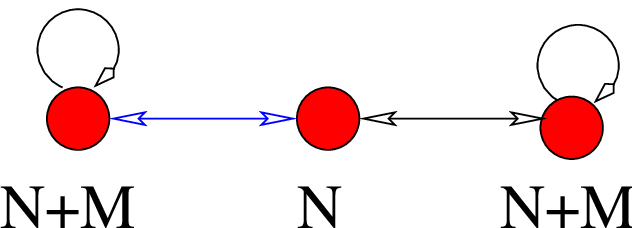}
\includegraphics[width=3cm]{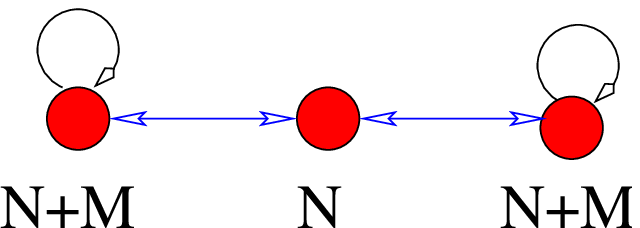}
\includegraphics[width=3cm]{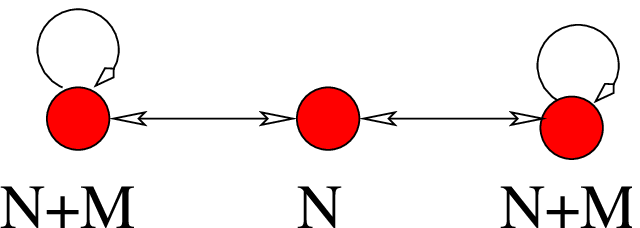}
\end{tabular}
\caption{Three different quivers from the deformed $L^{222}$. The
massive quarks are represented with blue lines, the massless quarks are
represented with black lines.}
\label{trecasi}
\end{center}
\end{figure}

We analyze here
the $L^{222}$ theory after a Seiberg duality.
The quiver of the complete theory (see figure \ref{l222quivero})
is related to the double conifold.
The superpotential is
\be
 W=-q_{21} X_{11} q_{12} + h q_{12} q_{23} q_{32} q_{21}
 -  q_{23} X_{33} q_{32}
 +q_{41} X_{11} q_{14} -  h q_{14} q_{43} q_{34} q_{41}
 +q_{43} X_{33} q_{34} 
\ee
and the geometry is given by the equation
\be
x^2 y^2 = wz \, .
\ee
We deform the geometry 
\be 
x(y-\xi_1)y(x-\xi_3)=wz \, . 
\ee 
This
deformation changes the constraints on the mesonic chiral ring. The
new constraints can be satisfied by adding in $W$ two linear terms
of the form $\xi_1 X_{11}$ and $\xi_3 X_{33}$, and
we have to switch on also two mass terms in the quarks fields.
Setting to zero one node, we can have the three different cases, as
shown in figure \ref{trecasi}.
They all have metastable vacua in the correct regime of couplings,
ranks and scales.

These models are similar to ISS, but with two differences: the
quartic term for the quarks and the mass term for some of the
quarks.

We study here the case with only one group of massive quarks (the
first case in the figure \ref{trecasi}), and then we comment on the
other at the end of this paragraph. A detailed analysis that
includes the three cases, for generic values of the masses of the
quarks, is in the Appendix \ref{luciano5}.

We choose the ranks of the groups to be
\be
N_2 = N \ \ \ \ N_1= N+M = N_3 \, .
\ee
The second node is treated as the gauge group and the other
two nodes as flavours.
The superpotential is
\be
W = -(\xi_1 X_{11} +\xi_3 X_{33}) -
q_{21} X_{11} q_{12} + h q_{12} q_{23} q_{32} q_{21}
-q_{23} X_{33} q_{32} +
h\xi_1 q_{32}q_{23} \, .
\ee

We then solve the equations of motion for the various fields,
recognizing the ISS rank condition, responsible for breaking of
supersymmetry. The $F$-terms fix the vacuum to be \be
q_{12}=q_{21}^T= \left(
\begin{array}{c}
\sqrt{\xi_1}\\
0
\end{array}
\right)
\ \ \ \
q_{32}=q_{23}^T
\left(
\begin{array}{c}
\sqrt{\xi_3}\\
0
\end{array}
\right)
\ \ \ \
X_{11} =
\left(
\begin{array}{cc}
0&0\\
0&\chi_1
\end{array}
\right)
\ \ \ \
X_{33} =
\left(
\begin{array}{cc}
\xi_1&0\\
0&\chi_3
\end{array}
\right) \, .
\ee 
In the Appendix \ref{luciano5} we show that this vacuum
is stable up to one loop corrections, fixing the pseudomoduli to
$\langle \chi_1 \rangle=0$ and $\langle \chi_3 \rangle=\xi_1$. The
two breaking sectors are separated at the one loop level, and their
quantum corrections are as in ISS. The $\xi_i$ deformations have
thus lead to supersymmetry breaking vacua.

The
strong dynamics of the
gauge group restores supersymmetry,
and is geometrically described by the
$\epsilon$ deformation
\be
\label{firsto}
(x-\xi_1) \left(x(y-\xi_3)-\epsilon\right)y = wz \, .
\ee

In the field theory analysis
we explore the large field region for the mesons, by integrating
out the massive fields, and by taking into account
the non perturbative contributions due to gaugino condensation.
The low energy superpotential results
\be
\label{irL222}
W_{IR} =N\left(\Lambda^{-2M-N}
\det{X_{11}}\det{X_{33}} \right)^{\frac{1}{N}}
-
\left(\xi_1 \Tr X_{11} + \xi_3 \Tr X_{33} \right) 
\ee
which guarantees the long life of the
vacuum\footnote{
The restoration of supersymmetry in the
other cases in figure \ref{trecasi} follows directly.}.

On the other hand,
we can use the geometric techniques
of Appendix \ref{secagana}
to recover the same low energy superpotential (\ref{irL222})
from the geometry (\ref{firsto}).
Relabeling the variables in (\ref{firsto}) by
$y = (x'-y')$ and $x = (x'+y')$ we can rewrite
\be
(x'-y'-\xi_1)
\left((x'-y')((x'+y'-\xi_3)-\epsilon\right) = wz \, .
\ee
The geometric superpotential is
\be\label{IRSPOT}
W_{IR}(S,X_{11},X_{33})=N_2 S \left (\log \frac{S}{\Lambda^3_m}-1 \right)
-\frac{t}{g}S + W_{adj}(S,X_{11}) + W_{adj}(S,X_{33}) \, .
\ee
The two contributions $W_{adj}$ derive from the singularities of the geometry.
Repeating the computations as in Appendix \ref{secagana} we have
\bea
W_{adj}(S,X_{11}) =&& \int \left(y'+\xi_1-\frac{\xi_3}{2}-\sqrt{\left(y'-
\frac{\xi_3}{2}\right)^2+\epsilon}\right)dy'
\nonumber \\
W_{adj}(S,X_{33}) =&& \int \left(\frac{\xi_3}{2}-\sqrt{\left(y'-\frac{\xi_3}{2}
\right)^2+\epsilon}+y'\right)dy' \, .
\eea
In the previous integral we identify $2S$ with $\epsilon$
and $y'$ with the vev of the adjoint fields $X_{11}$ and $X_{33}$ respectively.
In the regime
$y' \gg \epsilon, \xi_i$
we can
compute the integrals
expanding at first order in $\epsilon$ and $\xi_i$,
obtaining the superpotential for the interaction between the glueball field and the adjoint fields
\be\label{eqantonione}
W_{adj}(S,X_{11})+W_{adj}(S,X_{33})=
\xi_1 \Tr X_{11} + \xi_3 \Tr X_{33} -
S \log \det \left(\frac{X_{11}}{\Lambda_m}\right)
-
S \log \det \left(\frac{X_{33}}{\Lambda_m}\right) \, .
\ee
The equation for the glueball field $S$ can be derived from (\ref{IRSPOT})
and (\ref{eqantonione}). Solving for $S$ and ignoring the
multi-istanton contributions we have \be\label{formadiS} S =
\left(\Lambda_m^{3N_2-N_1-N_3}
\det{X_{11}}\det{X_{33}}\right)^{\frac{1}{N_2}} =
\left(\Lambda_m^{-N-2M}
\det{X_{11}}\det{X_{33}}\right)^{\frac{1}{N}} \, .
\ee 
Substitution of
(\ref{formadiS}) in (\ref{IRSPOT}) gives the same low energy
superpotential of field theory (\ref{irL222}), up to an overall
sign.

The $\epsilon$ deformation has lead to supersymmetry restoration.

\subsection{The $L^{333}$ theory}
\begin{figure}[h!!!]
\begin{center}
\includegraphics[width=5cm]{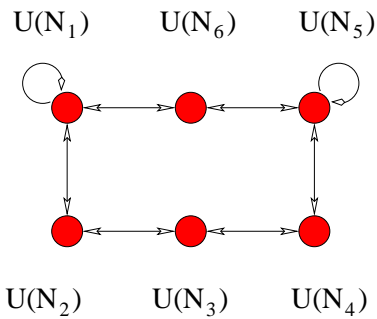}
\caption{The $L^{333}$ theory after a Seiberg duality on node $6$.}
\label{l333}
\end{center}
\end{figure}

Here we
search for metastable vacua in
an $L^{333}$ theory, after performing on it some Seiberg dualities.
This theory has six nodes without
adjoint fields,
with
superpotential
\be
W = \sum_{i=1}^{4} (-1)^i h q_{i,i+1}q_{i+1,i+2}q_{i+2,i+1}q_{i+1,i}
-h q_{56}q_{61}q_{16}q_{65}+ h q_{61}q_{12}q_{21}q_{16} \, .
\ee
A Seiberg duality on the sixth node
and integration out of the massive matter.
leads to the superpotential
\bea
W =&& - q_{61}X_{11}q_{16} + q_{21}X_{11}q_{12} - h q_{12}q_{23}q_{32}q_{21}
    + h q_{23}q_{34}q_{43}q_{32} - h q_{34}q_{45}q_{54}q_{43}
  \nonumber \\
&& + q_{45}X_{55}q_{54}-q_{65}X_{55}q_{56}
     + h q_{56}q_{61}q_{16}q_{65} 
\eea
with the quiver given in figure \ref{l333}.
The geometry is then deformed by the $\xi_i$ terms to 
\be 
x^2 y^2
(y-\xi_1)(x-\xi_5)=wz \, .
 \ee 
 This deformation corresponds in the field
theory to linear terms $\xi_1 X_{11}$ and $\xi_5 X_{55}$ in the
superpotential. For consistency with the $F$-term constraints, we
also add some mass term for the bifundamentals, i.e. 
\be 
\Delta W =
-\xi_1 X_{11}+\xi_5 X_{55}+h\xi_1 q_{23} q_{32}-h\xi_5 q_{43}q_{34} \, .
\ee 
We set the ranks of the groups as follows 
\be 
N_1=N_5=N+M \qquad
N_2=N_4=N \qquad N_3=N_6=0  \, .
\ee 
We then obtain two decoupled ISS like
models that break supersymmetry through rank conditions for the
mesons $X_{11}$ and $X_{55}$.

The supersymmetric vacua can be recovered by adding the non
perturbative contributions arising for each gauge group.
%
From the geometry, restoration of supersymmetry can be
described by the $\epsilon_i$ deformations
\be
x y (x(y-\xi_1)+\epsilon_1)((x-\xi_5)y-\epsilon_2)=wz \, .
\ee
This deformed geometry
gives, with the techniques of Appendix \ref{secagana},
the right
%
low
energy superpotential
that leads to
the supersymmetric vacua.

\subsection{Extension}
We now briefly outline a procedure
for finding metastable
vacua in a generic $L^{aaa}$ theory.

The strategy again consists in breaking the quiver into a set of
shorter quivers, each one metastable.

We study a phase of the theory, derived
by acting with Seiberg dualities,
which has a number of $a$ adjoint fields if
$a$ is even and $a-1$ if $a$ is odd.

We then set to zero the right
nodes\footnote{We set to zero only not consecutive nodes.}
in order to
obtain a set of decoupled theories that
have the same structure of the
deformed
$L^{222}$ and $L^{333}$
studied above.
This can be done choosing appropriately the Seiberg phases.

\begin{figure}[h!!!]
\begin{center}
\begin{tabular}{c}
\includegraphics[width=10cm]{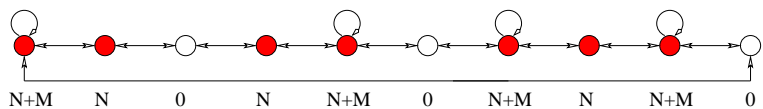}
\end{tabular}
\end{center}
\caption
{Quiver for the deformed $L^{555}$ theory.}
\label{figurmea2}
\end{figure}

We now show how to proceed in a simple example, $L^{555}$ in figure
\ref{figurmea2}.
We perform Seiberg dualities on the sixth and on the tenth node and
obtain $4$ adjoint fields.
We then deform the geometry in such a way that, in the field theory
description, all the adjoint fields get linear terms
\be
\label{geol555}
x^3 y^3 (y-\xi_1)(x-\xi_5)(y-\xi_7)(x-\xi_9) = wz \, .
\ee
Indeed, this deformation give rise to linear terms for all the adjoint
fields, and masses for some of the quarks.  The new superpotential
contribution is
\footnote{Other choices for the masses of the quarks are possible, and
all of them lead to metastability.}
%
\be 
\Delta W= \xi_1 q_{23}q_{32} +\xi_5 q_{34}q_{43}+\xi_7
q_{56}q_{65} +\xi_9 q_{110}q_{101}- \xi_1 X_{11} - \xi_5 X_{55}
-\xi_7 X_{77} -\xi_9 X_{99}  \, .
\ee 
We now set to zero the third, the
sixth and the tenth node. In this way we decompose the theory in
three different metastable sectors. The first two sectors have the
same structure of $L^{333}$, whereas the last sector is like the
theory emerging from a $L^{222}$. In short we have decomposed the
$L^{555}$ as $L^{222}$ and $L^{333}$.

Supersymmetry restoration is achieved
in each sector separately.
From the geometric point of view this transition is read as an
$\epsilon$
deformation of (\ref{geol555}) to
\be
x^2 y^2
((y-\xi_1)x-\epsilon_1)(y(x-\xi_5)-\epsilon_2)((y-\xi_7)(x-\xi_9)-\epsilon_3)
= w z
\ee
where the three $\epsilon_i$ take into account the deformations on the
moduli space imposed by the strong dynamics of the three groups
that we considered as gauge groups.
Using the geometric techniques
of Appendix \ref{secagana}
it is possible also in this
case to recover the correct low energy behavior
in the supersymmetric vacua.
The three different deformation parameters $\epsilon_i$
are interpreted
as the three glueball fields of the three gauge groups.
%

\subsection{Back to $L^{aba}$}

Up to now we have found metastable vacua in all the $L^{aaa}$
theories (with $a>1$) and in $L^{aba}$ with the constraint $b-a>a$.
The study of the $L^{aaa}$ theories gives us a way out from the
constraints imposed on  $L^{aba}$.
If we have an $L^{aba}$ theory with $b-a<a$ we
have indeed to look for a different Seiberg phase. 
Given an $L^{aba}$ theory one can find a dual theory with
at most
$b+a-2$ adjoint fields, instead of $b-a$.

\begin{figure}[h!!!]
 \begin{center}
\includegraphics[width=10cm]{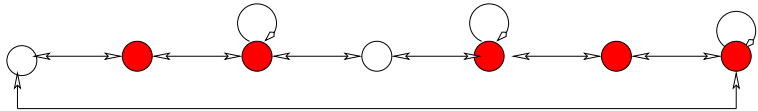}
\caption{$L^{343}$ theory with a Seiberg duality on the fourth node.}
  \label{L343}
\end{center}
\end{figure}

We proceed in a simple example: the $L^{343}$ theory. A Seiberg
duality on the fourth node gives the quiver in figure \ref{L343}.
By adding a linear deformation for each adjoint field, the geometry
becomes 
\be 
x^2 y^2 (y-\xi_1) (x-\xi_2) (y-\xi_3) =wz \, .
\ee 
We can now
set some node to zero and obtain a set of decoupled theories with a
metastable IR behavior. A possible choice is shown in figure
\ref{L343}, where we set to zero the white nodes. We have broken up
$L^{343}$ theory in two sectors: the first one has the same property
of metastability of $L^{121}$, and the second one of $L^{222}$, the
double conifold.

Supersymmetry is restored by the geometric $\epsilon$ deformation
\footnote{Note that in this case we chose massless quarks in the
last two lines of the quiver.}
\be
x y ((y-\xi_1)x-\epsilon_1) (x-\xi_2)  (xy-\epsilon_2) (y-\xi_3) =wz
\ee
through the strong dynamics
of the gauge group in each decoupled sector.

\section{Beyond the $L^{aba}$ cases}\label{metgen}

In the previous sections we performed metastable supersymmetry
breaking in the family of $L^{aba}$ singularities. An immediate
generalization is the embedding of $L^{aba}$ in larger singularities
and the recovering of metastable dynamics in the IR.

We need to start with a UV quiver gauge theory and
flow by way of the renormalization group
to a set of gauge theories
with fewer gauge groups.
These theories are decoupled at low energy,
and they keep at least one $L^{aba}$
singularity.
These singularities trigger metastability in the IR.

In the RG flow to the IR two different decouplings are possible,
the resolution and the deformation of the mother singularity.
\\
Blowing up two spheres gives first the resolution of the mother
singularity. The daughters singularities are geometrically separated
by the volume of these two spheres. This corresponds to the motion
in the Kahler moduli space of the singularities.
\\
The second one, the deformation, is achieved by blowing-up three spheres.
Here the singularities are separated by the volume
of the three spheres.
\\
In both cases the IR theories decouple at the level of massless
states and the masses of the messenger fields are controlled by the
volume of the two and three spheres respectively.
\\
We now describe these two possibilities by
proceeding with pictures and examples.

\begin{figure}[h!!!]
\begin{center}
\includegraphics[scale=0.55]{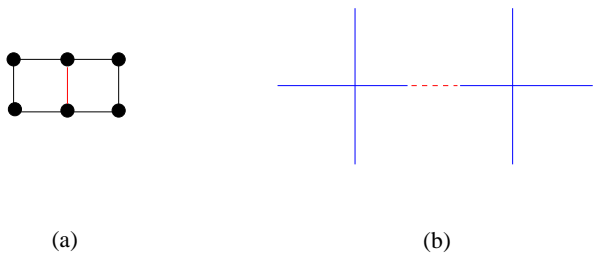}
\caption{The toric resolution of the double conifold: $L^{222}$. (a)
The toric diagram representation, (b) the dual diagram: the broken red
arrow parametrize the volume of the blown up two sphere.}
\label{conr}
\end{center}
\end{figure}

The graphical resolution
of a singularity in
the toric language
corresponds to drawing a line in the toric diagram (the red line in
our figures) and a perpendicular line in the dual diagram (the
dashed line). This last line parametrizes the volume of the two
sphere (see Figure \ref{conr}).

\begin{figure}[h!!!]
\begin{center}
\includegraphics[scale=0.55]{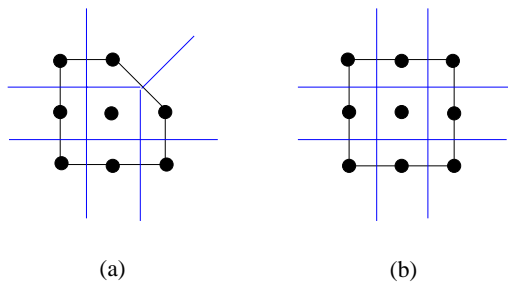}
\caption{The toric diagrams and the dual diagrams for (a) $PdP_4$ and
(b) $PdP_5$.}
\label{PdP}
\end{center}
\end{figure}

A natural laboratory for these constructions is
the family of Pseudo del Pezzo singularities $PdP_n$. These are
complex cones over $\mathbb{P}^2$, blown up at $n$ non-generic
points. This blowing up generates lines of
singularities passing trough the tip of the cones (Figure
\ref{PdP}).

\begin{figure}[h!!!]
\begin{center}
\includegraphics[scale=0.55]{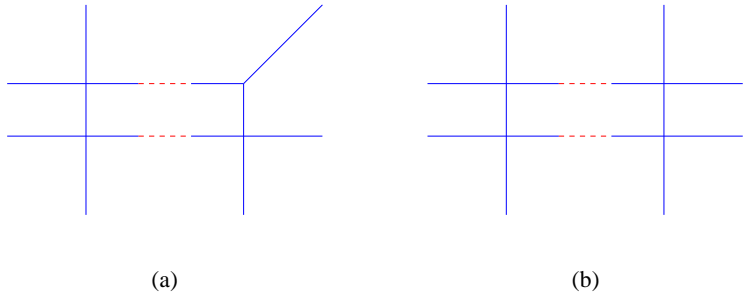}
\caption{Resolutions of (a) $PdP_4$ and (b) $PdP_5$.}
\label{PdPres}
\end{center}
\end{figure}

In the $PdP_4$ and $PdP_5$ singularities it is possible to recover
two of the singularities that show a metastable behavior, $L^{121}$
($SPP$) and $L^{222}$ (double conifold), through the resolution of
the singularities as shown in Figure \ref{PdPres}.
We first assign a set of fractional branes to the mother singularity
such that
it reproduces, at least for one of the daughter singularities, the
set of fractional branes that has metastable non
supersymmetric vacua.
We turn then on Kahler moduli deformations, decoupling in the IR one
$L^{121}$ and one $L^{222}$ singularities from the $PdP_4$. For the
$PdP_5$ singularity we can decouple two $L^{222}$ singularities.  In
each situation the two decoupled IR theories are separated at the
level of massless states \footnote{As discussed in
\cite{GarciaEtxebarria:2006rw} using Kahler
  moduli space deformations it is possible to compute the mass of the
  ``messenger particles'' but the Kahler moduli remain  free
  parameters to be stabilized in some way.}.
Finally, metastable supersymmetry breaking can be realized, since we
can deform the $A_1$ singularities belonging to one or to both the
IR theories.
\begin{figure}[h!!!]
\begin{center}
\includegraphics[scale=0.55]{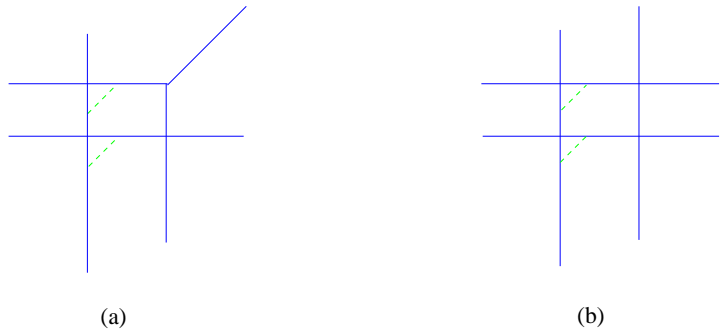}
\caption{Deformations of (a) $PdP_4$ and (b) $PdP_5$.}
\label{PdPdef}
\end{center}
\end{figure}

The other possibility for the decoupling of a mother singularity is
the deformation (see section \ref{labasing} for a graphical
description).
It furnishes a second embedding of $L^{121}$ and $L^{222}$ into $PdP_4$ and
$PdP_5$.
These configurations are described in Figure \ref{PdPdef}.

We have to distribute the fractional branes at the mother
singularity in such a way that they lead to the complex moduli
deformation. Gaugino condensation is then induced by the strong
dynamics of some gauge groups. This decoupling leads to the
remaining daughter singularities in the IR,
and, in this case, we are left with
$L^{121}$ and $L^{222}$.
We can move in the complex moduli space deformations of the non
local singularity, reproducing the supersymmetry breaking behaviour
of the $L^{aba}$ theories.
\\
The advantage of this procedure is that the
moduli associated with the volumes of the three spheres are
automatically stabilized by the strong IR gauge dynamics.
The drawback
is that the computation of the masses of the
messenger sector is not straightforward.
\\
\\
Following the two procedures explained in this section
and the methods developed in
\cite{GarciaEtxebarria:2006rw} many examples, useful for model building,
can be studied.

There exist conical singularities that provide
extensions of MSSM as the IR limit of the dynamics of D3 branes
put at the tip of the cone.
The easiest example
is given by
$D3$ branes at $dP_0$ singularity.

Here, by using either Kahler moduli deformations
or complex moduli deformations,
it is possible to separate a singularity into
a $dP_0$ sector and some $L^{aba}$ sector.
In the IR, $dP_0$ is an extension of the MSSM, $L^{aba}$ is the hidden
supersymmetry breaking sector, and the massive fields are the messengers.
It is possible to find many examples of singularities that, after the
resolution, decouple in a MSSM like sector and in a hidden
supersymmetry breaking sector, also metastable.  We show here two
possibilities.

\begin{figure}[h!!!]
\begin{center}
\includegraphics[scale=0.5]{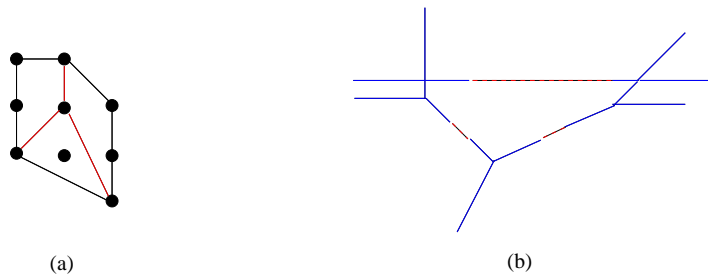}
\caption{The resolved toric diagram (a) and the dual diagram (b). The
triangle at the bottom is the $dP_0$ singularity that represents
the ``visible sector'', the polygon on the top are two decoupled SPP
singularities that represent the supersymmetry breaking sector.}
\label{dP0res}
\end{center}
\end{figure}

\begin{figure}[h!!!]
\begin{center}
  \includegraphics[scale=0.5]{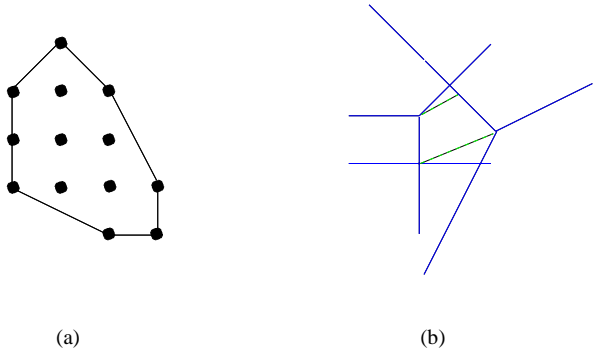}
\caption{(a) The toric diagram of the mother singularity and (b) the
deformed dual diagram that contain the $dP_0$ visible sector and
the $SPP$ supersymmetry breaking sector.}
\label{dP0def}
\end{center}
\end{figure}

The first one, in Figure \ref{dP0res},
admits a resolution that decouples in the IR
a $dP_0$ and two $SPP$ singularities.
The $dP_0$
plays the role of phenomenological sector,
while the two
$SPP$ singularities play the role of
supersymmetry breaking hidden sectors.
The second one, in figure \ref{dP0def},
admits a complex deformation. It decouples a $dP_0$ sector
and a single $SPP$ sector.
\clearpage

\addcontentsline{toc}{section}{Conclusions}
\section*{Conclusions}

In this paper we discussed 
the geometric interpretation of metastable vacua for systems of D3
branes at non isolated deformable toric CY singularities. We have
generalized the analysis done in \cite{Buican:2007is} to the
infinite family of $L^{aba}$ singularities and we have proposed the
embedding of these theories in bigger singularities.

The dynamical generation of the $\xi$ deformation which sets the
scale of the supersymmetry breaking is still an open problem. Since
much is known about the metric of the $L^{aba}$ spaces, another
challenging question regards metastability in the gauge/gravity
correspondence. The models here studied may play the role of hidden
sector in mechanisms of gauge mediation of supersymmetry breaking
\cite{Dine:1981gu} in metastable vacua \cite{Dine:2006xt}.

\section*{Acknowledgments}

It is a pleasure to thank Loriano Bonora, Inaki Garcia-Exteberria,
Angel Uranga and Alberto Zaffaroni for many enlightening discussions.
We are also grateful to Riccardo Argurio, Radu Tatar and Sean McReynolds
for useful discussions and comments.
A.~A.~, L.~G.~ and A.~M.~
are supported in part by INFN, PRIN prot.2005024045-002 and the
European Commission RTN program MRTN-CT-2004-005104.
A.M. is also supported by Fondazione Angelo Della Riccia.
D.~F.~ is supported in part by INFN and the
Marie Curie fellowship under the program EUROTHEPHY-2007-1.

\appendix

\section{Toric diagrams}
\label{APPTORICA}

From the algebraic-geometric point of view  the data of a conical
toric Calabi-Yau are encoded in a rational polyhedral cone
$\mathcal{C}$ in $\mathbb{Z}^3$ defined by a set of vectors
$V_{\alpha}$ $\alpha=1,...,d$. For a CY cone, using an $SL(3,
\mathbb{Z})$ transformation, it is always possible to carry these
vectors to the form $V_{\alpha}=(x_{\alpha},y_{\alpha},1)$. In this
way the toric diagram can be drawn in the $x,y$ plane (see for
example Figure \ref{tqspp}). The CY equations can be reconstructed
from this set of combinatorial data using the dual cone
$\mathcal{C}^*$.
\\
The two cones are related as follow. The geometric generators for
the cone $\mathcal{C}^*$, which are vectors aligned along the edges
of $\mathcal{C}^*$, are the perpendicular vectors to the facets of
$\mathcal{C}$.  To give an algebraic-geometric description of the
CY, we consider the cone $\mathcal{C}^*$ as a semi-group and find
its generators over the integer numbers. The primitive vectors
pointing along the edges generate the cone over the real numbers but
we generically need to add other vectors to obtain a basis over the
integers. Denote by $W_j$ with $j=1,...,k$ a set of generators of
$\mathcal{C}^*$ over the integers.  To every vector $W_j$ one can
associate a coordinate $x_j$ in some ambient space. $k$ vectors in
$\mathbb{Z}^3$ are linearly dependent for $k > 3$, and the additive
relations satisfied by the generators $W_j$ translate into a set of
multiplicative relations among the coordinates $x_j$. These are the
algebraic equations defining the six-dimensional CY cone.
\\
All the relations between points in the dual cone become relations
among mesons in the field theory. In fact, there exists a one to one
correspondence among the integer points inside $\mathcal{C}^*$ and
the mesonic operators in the dual field theory, modulo F-term
constraints \footnote{For the relations between the chiral ring of
toric CFT and the geometry of the singularities see
\cite{Hanany:2006nm,Butti:2006hc,Butti:2006nk,Benvenuti:2006qr,
Butti:2006au}.}. To every integer point $m_j$ in $\mathcal{C}^*$ we
indeed associate a meson $M_{m_j}$ in the gauge theory with $U(1)^3$
charge $m_j$, which uniquely determine them. The first two
coordinates $ Q^{m_j}=(m_j^1,m_j^2)$ of the vector $m_j$ are the
charges of the meson under the two flavour $U(1)$ symmetries.  Since
the cone $\mathcal{C}^*$ is generated as a semi-group by the vectors
$W_j$ the generic meson will be obtained as a product of basic
mesons $M_{W_j}$, and we can restrict to these generators for all
our purposes.  The multiplicative relations satisfied by the
coordinates $x_j$ become a set of multiplicative relations among the
mesonic operators $M_{W_j}$ inside the chiral ring of the gauge
theory. It is possible to prove that these relations are a
consequence of the F-term constraints of the gauge theory.  The
abelian version of this set of relations is just the set of
algebraic equations defining the CY variety as embedded in
$\mathbb{C}^k$.  In the example of SPP from the four mesons
$x,y,z,w$ we associate the quadric $xy^2=zw$ in $\mathbb{C}^4$.

\section{The ISS theory}
\label{ISSmodello}
The existence of
long living metastable vacua seems to be a rather generic phenomenon in
$\mathcal{N}=1$ supersymmetric gauge theory.
Their existence has been shown in simple theories, like
SQCD with massive flavors \cite{Intriligator:2006dd}.
%
Consider a $SU(N_c)$ gauge theory with $N_f$ fundamental massive
flavors and superpotential \be W =m_Q Q \tilde Q \ee in the free
magnetic phase, when $N_c<N_f<\frac{3}{2}N_c$. The theory is UV
free, since the beta function $b=3N_c-N_f$ is positive. In order to
analyze the low energy dynamics and explore supersymmetry breaking,
we need a weakly coupled description of this theory, where
perturbative techniques can be used. This is achieved by performing
a Seiberg duality at the strong coupling scale of the UV theory.

The Seiberg dual magnetic theory has gauge group
%
$SU(N_f-N_c)$, the same
flavour symmetry, and
superpotential
\be 
W = h M q \tilde q - h \mu^2 M 
\ee 
where the $q$ fields are the
magnetic quarks, and the meson $M$ is an elementary field, which
corresponds, up to rescaling, to the electric gauge singlet $Q
\tilde Q$.  This theory can be studied perturbatively, since $b =
2N_f-3N_c$ is now negative.

Not all the $F$-equations for
the $M$ field can be solved. This breaking condition has been called
Rank condition, since it is due to the fact that the meson $M$ gives a
squared matrix $\delta_i^j$ of rank $N_f$, while the other squared
matrix involved in the equation ($q_i^\alpha \tilde q_\alpha^j$) has
rank $N_f-N_c$.
Hence 
there are $N_c$ equations that cannot
be solved, breaking supersymmetry, and giving a non zero value to the
scalar potential.


The $F$ and $D$ equations of motion
fix the vev of the fields
in the tree level supersymmetry breaking vacuum
to be
\be
\label{ISSvevalbe}
q =
\left(
\begin{array}{c}
\mu e^{\theta} \mathbf 1_{N_c}\\
0
\end{array}
\right)
\ \ \ \ \ \
q =
\left(
\begin{array}{c}
\mu e^{-\theta} \mathbf 1_{N_c}\\
0
\end{array}
\right)
\ \ \ \ \ \ \
 M = \left(
\begin{array}{cc}
0&0\\
0&\chi
\end{array}
\right) \, .
\ee
Not all the directions are lifted at the classical level,
and some pseudo-flat directions can destabilize this
tree level vacuum.
Indeed
the $\theta$ and $\chi$
fields are
 pseudo Goldstones, i.e. flat directions
not associated to any broken
global symmetries, and not protected at the quantum level.
%
Precisely, analyzing
the fluctuations
around the vacuum (\ref{ISSvevalbe}) using only the $F$-term
contributions,
other flat directions arise in the upper part
of the magnetic quarks.
However these directions are lifted
by the $D$-term contribution to the scalar
potential for the gauge group $SU(N_f-N_c)$.
%

Hence the potentially dangerous flat directions are the
$\theta+\theta^{*}$ and $\chi$ fields.
%
%
Their stability has been checked \cite{Intriligator:2006dd} at one loop
using the Coleman-Weinberg effective potential.
It has been shown that,
at one loop,
they acquire positive mass squared,
and the minimum is fixed in
\be
\theta+\theta^*=0 \ \ \ \ \ \chi=0 \, .
\ee

In the analysis of this paper
the $\theta+\theta^*$ pseudomodulus does not appear,
since we study theories with a $U(N)$
and not $SU(N)$.
In the $U(N)$  case
there is
a further contribution
from the $D$-terms (the trace) to the scalar potential,
which stabilize the $\theta+\theta^*$ fields
at the tree level.\\

A relevant aspect for the non supersymmetric vacuum is
the estimation of its lifetime.
In fact since SQCD with massive flavours
 has Witten index $N_c$
one expects to have $N_c$ supersymmetric vacua
elsewhere in the field space.
Thus we have to check that the non supersymmetric vacuum
has a low decay rate into the supersymmetric one.

The supersymmetric vacua can be found \cite{Intriligator:2006dd}
by taking into account also the
gaugino condensation contribution to the
superpotential
\be
W_{\text{dyn}}=N_c \left(\Lambda^{3N_c-N_f} \det{M}
\right)^{1/N_c} \, .
\ee
Now we can solve the equation
of motion finding zero vev for the quarks and
\be
\langle M \rangle = \frac{\mu}{h}
\left(\frac{\Lambda_m}{\mu}\right)^{\frac{N_f-3N_c}{N_f-N_c}} \, .
\ee
These supersymmetric vacua are
parametrically far from the non supersymmetric one,
and this guarantees the long lifetime of
the non supersymmetric vacuum.

\subsection{ISS like models with gauged flavour}
\label{gaugeunimportant}
In the main text we look for
ISS like vacua in quiver gauge theories.
The main difference between SQCD and these theories
is that in the latter the symmetries
are all gauged, and hence also the
flavour groups are gauged
as well.
%
%
In the analysis of the moduli spaces
the gauge contributions of these
groups may become relevant.

Such groups may
develops a strong dynamics that ruins the conclusions about the
lifetime of the metastable vacua, since new supersymmetric vacua arise.

Another problem is that
%
some fields
charged under these groups could take non zero vev in the
meta-stable vacua. This makes the one loop computation
difficult, since we should
take into account the $D$-term corrections to the effective
potential.
In fact the mass matrices
which appear in the Coleman Weinberg
potential are built using the $F$-terms of
the superpotential, and the $D$-terms arising from the gauge groups.
The $D$-term contributions to the mass matrix are irrelevant
with respect to the $F$-term ones only if
the corresponding gauge group is very
weakly coupled.
%

The problems associated with the gauging of the
flavour symmetries has already been handled in
\cite{Argurio:2006ny, Buican:2007is,Amariti:2007am,Forste:2006zc}
with different solutions.
Basically one needs a scheme where the gauge contributions of such groups can
be ignored.
If these groups are IR free
in the Seiberg dual description, the way out consists of tuning their
Landau pole to be much higher than the Landau pole $\Lambda_m$ of the
dualized gauge group.
In the opposite case, 
the gauged flavour groups are UV free.
In this case we have to choose the opposite tuning,
i.e.  their strong coupling scale must be much lower than $\Lambda_m$ and also
lower than the supersymmetry breaking scale.  Such tunings make
the gauge contributions of the flavour groups 
negligible, and the problems mentioned above are avoided.

\section{Geometric transition and the superpotential}
\label{secagana}
In this Appendix we review the
geometric transition techniques of \cite{Aganagic:2007py}
for computing the low energy superpotential
from the geometrical data.
The computation is
illustrated here for the $\epsilon$-deformed geometries.
These deformations are due to
the strong dynamics developed by the gauge groups that
lead to the supersymmetric vacua.

With this technique it is possible to write the
superpotential for the gaugino condensate and its interaction
with the adjoint fields, which are the mesons describing the low energy
theory.
The dynamical deformation $\epsilon$ of the geometry is
related to the gaugino condensate, while the adjoint field is interpreted
as the location of the $D5$-branes relative to the dynamically deformed
conifold.

In the SPP example, the deformed geometry is
\be\label{condeadef}
  \left(x(y-\xi)-\epsilon \right)y=wz
\ee
and the glueball field is given by $\epsilon=2 S$.

The low energy superpotential  $W_{IR}$
is composed by two contributions
\be
\label{WIRalbeapp}
W_{IR}=W_{GVW}(S)+ W_{adj}(S,X)
\ee
the first one involves the glueball
field $S$ whereas the second one is the
contribution of the adjoint field $X$.

The superpotential for the glueball field is the
GVW
flux superpotential
\be \label{GVWW}
W_{GVW}(S)=\int H \wedge \Omega
=N S \left( \log {\frac{S}{\Lambda_m^3}}-1\right)
+\frac{t}{g_s} S \, .
\ee
This perturbative superpotential is a function of the
glueball field $S$ and of a parameter $t$.
The $t$ parameter takes into account the multistanton
contribution to the low energy superpotential.
In fact since we
have D5-branes wrapping rigid $\mathbb{P}^1$ in a Calabi-Yau,
D1-brane istantons wrapping the $\mathbb{P}^1$ generate a
superpotential proportional to $\exp^{-\frac{t}{g_s N}}$
with $t=\int_{S^2}B^{NS}+i g_sB^{RR}$.
Expanding
with respect of $t$ in the low energy theory
we can take into account the multistanton
contribution.


In \cite{Aganagic:2007py} it has been shown
how to compute from
geometrical data the adjoint contribution $W_{adj}(S,X)$
to
the low energy superpotential.
%
%
It is given by the
integral over holomorphic 3-form
\be
W_{adj}(S,X)=\int_{\Gamma}\Omega
\ee
where $\Gamma$ is a 3-chain bounded by the 2-cycle that the D5 brane wraps.
This can be computed writing the
geometry (\ref{condeadef})
in terms of new variables $x=x'-y'$ and
$y=x'+y'$
\be
\prod_{i=1}^{3}(x'-{x'}_i(y'))=wz
\ee
and evaluating
\be
\label{SPPalbeapp}
W_{adj} = \int ({x'}_3(y')-{x'}_1(y'))dy' \, .
\ee

More generally,
\cite{Aganagic:2007py} if we have a
geometry of the form
\be
\prod_{i=1}^{n}(x'-{x'}_i(y'))=wz
\ee
the contribution of the $j$-th node to this superpotential is
of the form
\be
W_{j,adj} = \int ({x'}_j(y')-{x'}_{j+1}(y'))dy' \, .
\ee

In the SPP case the only node in the quiver with the
adjoint field is $N_1$, and
indeed the contribution to the superpotential is
(\ref{SPPalbeapp}).
%
%
In the regime where all the deformations
are lower than $y'$ ($y' \gg \epsilon, \xi_i$),
we can expand the
integral
(\ref{SPPalbeapp})
at first order in $\epsilon$, and obtain
\be
\label{unasola}
W_{adj} = \xi  X_{11} -S  \log\left( \frac{X_{11}}{\Lambda_m}\right)
\ee
where we have identified $\epsilon=2S$.
From the full low energy superpotential $W_{IR}$ (\ref{WIRalbeapp})
we can now obtain
a description in terms of the adjoint field only. This is
achieved by integrating out the
glueball field, using
$(N+M)$ copies of (\ref{unasola})
\be
S =
\left(\Lambda_m^{2N-M} \det X_{11} \right)^{1/N} \text{e}^{-t/g_s}
\sim \left( \Lambda_m^{2N-M} \det X_{11} \right)^{1/N}
\ee
without considering multi-istanton contributions.
With this procedure we recover the
expected result
\be
W_{IR}  = \xi_1 X_{11} - N \left( \Lambda_m^{2N-M} \det X_{11} \right)^{1/N}
\ee
which is understood in field theory as the
low energy contribution to the superpotential
due to the gaugino condensation of the node $N_1$.

\section{Details on the non supersymmetric vacua}
\label{appB}
In this Appendix we discuss the stability of the
non supersymmetric vacua studied in the rest of the paper.
The relevant aspects in the analysis of metastable vacua
are related to the tree level flat directions that can
arise in the scalar potential around the would be minimum.
%
If these directions are
not 
related to any broken global symmetry they are pseudomoduli,
and they
have to be lifted classically or quantum mechanically.
Even if these directions arise in a
sector which is supersymmetric up to the third order
in the fluctuations around the vacuum,
we have to check that all of them
acquire positive squared masses.
Otherwise these fields can
acquire tachyonic masses due to their coupling
to the non supersymmetric sector
at higher order.
In the analysis we treat all the gauge groups as $U(n)$.
This implies that the $D$-term scalar potential for the fluctuations
around the minimum receives contributions not only from the $SU(n)$
part of the gauge groups but also from the $U(1)$'s.
These contributions could be relevant in some examples to
lift flat directions. We comment on this when needed.


A last comment 
is necessary. In the text
we called the complex deformations that lead to supersymmetry
braking $\xi_i$. In this Appendix we use a different notation,
denoting $\mu_i^2$ these deformations. In this way we
work with couplings of mass dimension one.


\subsection{$L^{131}$}
\label{luciano1}
\begin{figure}[h!!!]
\begin{center}
\includegraphics[width=4cm]{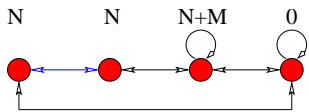}
\caption{The $L^{131}$ theory with $N_4=0$.
The blue line indicate the massive
fields}
\label{Appendix1}
\end{center}
\end{figure}

We analyze the quiver gauge theory
of figure \ref{Appendix1}
with superpotential
\be
W=X_{33} q_{32} q_{23}-\mu_3^2 X_{33} - h q_{12} q_{23} q_{32} q_{21}
+h m^2 q_{12}q_{21}
\ee
with $m^2 = \mu_3^2+\mu_4^2$.
The adjoint field has a linear term and
the quarks have a mass generally different from the
deformation of the adjoint field.
We take the ranks of the gauge groups as
\be
N_3=N+M \qquad N_2=N_3=N \quad N_4=0
\ee
with $M>N$. With this choice we are guaranteed
that the second node is infrared free.
We consider the other
groups less coupled.

Solving the equation of motion and
expanding around the tree level minimum
we have
\be
q_{32}=
\left(
\begin{array}{c}
\mu_3+ \sigma_1 \\
\phi_1
\end{array}
\right)
\quad
q_{23}
\left(
\begin{array}{c c}
\mu_3+\sigma_2 & \phi_2
\end{array}
\right)
\quad
X_{33}=
\left(
\begin{array}{c c}
\sigma_3 & \phi_3 \\
\phi_4 & \chi
\end{array}
\right)
\quad
q_{21}=\sigma_4
\quad q_{12}= \sigma_5
\ee
where $\chi$ is a classical flat direction
not associated to any broken symmetry.
The case with $\mu_4=0$ (and hence $m^2=\mu_3^2$)
is problematic since
in this case the quarks $q_{12}$  and $q_{21}$
are potentially dangerous tree level flat directions.

Now, the non supersymmetric sector (the fields $\phi_i$)
gives the usual O'Raifeartaigh like model
of ISS which gives positive squared mass through 1 loop corrections
to the
pseudomoduli\footnote{If the $U(1)$ factor of $U(N_2)$ decouples there is another pseudomodulus,
$\theta+\theta^*$, stabilized by 1-loop corrections (see Appendix \ref{ISSmodello}).}
 $\chi$.
%
The fields $\phi_i$ get tree level masses except the
Goldstone bosons as in the ISS model.

In the supersymmetric sector, the $\sigma_1,\sigma_2,\sigma_3$
fields are stabilized as in ISS. The fields $\sigma_4$ and $\sigma_5$
get non trivial squared mass $\sim |h m^2 - h \mu^2|^2=|h \mu_4^2|^2$.

\subsection{$L^{141}$}
\label{luciano2}
\begin{figure}[h!!!]
\begin{center}
  \includegraphics[width=5cm]{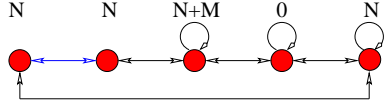}
\caption{The $L^{141}$ theory with $N_4=0$. The blue line indicate the massive
fields}
\label{Appendix2}
\end{center}
\end{figure}

We analyze here a more complicated example, explained in section
\ref{secl141}, that arises setting to zero a node in the $L^{141}$
quiver gauge theory.  The resulting quiver is reported in figure
\ref{Appendix2}
and the superpotential is the following
\be
W=h q_{12} q_{23} q_{32} q_{21} - \mu_3^2 X_{33}- X_{33} q_{32} q_{23}
+ \mu_5^2 X_{55} + X_{55} q_{51} q_{15}- h q_{51} q_{12} q_{21} q_{15}
+h m^2 q_{12}q_{21}
\ee
where all the adjoint fields receive a linear term.
From the geometric description we know that
\be
m^2= \mu_3^2+\mu_4^2-\mu_5^2
\ee
where $\mu_4^2$ is related to the node we have set to zero.
Having set the ranks of the gauge group to be
\be
N_1=N_2=N_5=N \qquad N_3=N+M \quad N_4=0
\ee
a rank condition mechanism is realized
for the $X_{33}$ meson.

Solving the equation of motion and
expanding around the tree level minimum
we have
\bea
&&q_{23}=
\left(
\begin{array}{c}
\mu_3+ \sigma_1 \\
\phi_1
\end{array}
\right)
\quad
q_{32}
\left(
\begin{array}{c c}
\mu_3+\sigma_2 & \phi_2
\end{array}
\right)
\quad
X_{33}=
\left(
\begin{array}{c c}
\sigma_3 & \phi_3 \\
\phi_4 & \chi
\end{array}
\right)
\non
\\
&&
q_{12}=\sigma_4
\quad q_{21}=\sigma_5
\quad
q_{51}=  \mu_5+\sigma_6
\quad
q_{15}= \mu_5+\sigma_7
\quad
X_{55}=\sigma_8 \, .
\eea
The non supersymmetric sector (the $\phi_i$ fields) is like the ISS model,
and give raise to
an O'Raifeartaigh model which stabilize at one loop the
pseudomodulus at $\chi=0$.

The supersymmetric sector (the $\sigma_i$ fields)
has the following superpotential at the relevant order
for the  mass matrix
\be
W=\mu_3 \sigma_3 (\sigma_1+\sigma_2)-h \mu_4^2 \sigma_4 \sigma_5
-  \mu_5 \sigma_8 ( \sigma_6+\sigma_7) \, .
\ee
The $\sigma_1, \sigma_2,\sigma_3$ fields behave exactly as in ISS:
some of them acquire tree level positive mass. The massless ones
are either Goldstone bosons either pseudomoduli. The latter
are lifted by the $D$ term potential for the $U(N_2)$ gauge group.

The $\sigma_4, \sigma_5$ fields have tree level masses
and this is due to the fact that we have turned on
all the possible deformation for the geometry, i.e. $\mu_4 \neq 0$.
Otherwise they would be dangerous flat directions.

The $\sigma_6, \sigma_7,\sigma_8$ fields
behave as the $\sigma_1, \sigma_2, \sigma_3$ sector.
However we note that here the pseudomoduli arising in these
fields are lifted by the $D$ terms of the $U(N_5)$ gauge group,
that we have considered less coupled than the
gauge group $U(N_2)$.
%
%

\subsection{$L^{151}$}
\label{luciano3}
\begin{figure}[h!!!]
\begin{center}
  \includegraphics[width=6cm]{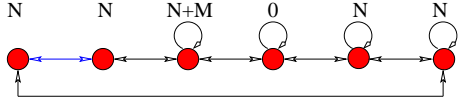}
\caption{The $L^{151}$ theory with $N_4=0$. The blue line indicate the massive
fields}
\label{Appendix3}
\end{center}
\end{figure}

We study here the quiver gauge theory presented in section \ref{secl151}.
The aim is to find the relevant aspects for the
generalization to the $L^{1n1}$ theory.
After setting to zero a node in the $L^{151}$ theory
we obtain the quiver in figure \ref{Appendix3}
with superpotential
\bea
W&=&X_{33} q_{23}q_{32}- h q_{12}q_{23}q_{32}q_{21}+
h q_{61}q_{12}q_{21}q_{16}-
X_{55}q_{56}q_{65}+X_{66}q_{65}q_{56}-X_{66}q_{61}q_{16}\non \\
&&-\mu_3^2 X_{33}+\mu_5^2 X_{55}+\mu_6^2 X_{66} + h m^2 q_{12}q_{21} \, .
\eea
The geometric description implies
\be
m^2=\mu_3^2+\mu_4^2-\mu_5^2-\mu_6^2
\ee
where the parameter $\mu_4$ is related to the
deformation for the node we have set to zero.
The ranks of the groups are taken to be
\be
N_3=N+M \qquad N_1=N_2=N_5=N_6=N \quad N_4=0 \, .
\ee
Solving the equation of motion and
expanding around the tree level minimum
we have
\bea
&&q_{23}=
\left(
\begin{array}{c}
\mu_3+ \sigma_1 \\
\phi_1
\end{array}
\right)
\quad
q_{32}
\left(
\begin{array}{c c}
\mu_3+\sigma_2 & \phi_2
\end{array}
\right)
\quad
X_{33}=
\left(
\begin{array}{c c}
\sigma_3 & \phi_3 \\
\phi_4 & \chi
\end{array}
\right)
\non
\\
&&
q_{12}=\sigma_4
\quad q_{21}=\sigma_5
\quad
q_{16}= \sqrt{\mu_5^2+\mu_6^2}+\sigma_6
\quad
q_{61}= \sqrt{\mu_5^2+\mu_6^2}+\sigma_7 \\
&&
X_{66}=\sigma_8
\quad
q_{65}= \mu_5+\sigma_9
\quad
q_{56}= \mu_5+\sigma_{10}
\quad
X_{55}=\sigma_{11} \, .
\eea
The non supersymmetric sector works as in the previous examples
and stabilize the pseudomodulus $\chi$ at $\chi=0$.
The supersymmetric sector (the $\sigma_i$) has, at the relevant order
for the mass matrix, the following superpotential
\be
W=\mu_6 (\sigma_{11}-\sigma_8)(\sigma_9+\sigma_{10})+
\sqrt{\mu_6^2+\mu_5^2} \, \sigma_8 (\sigma_6+\sigma_7)
- h \mu_4^2 \sigma_4 \sigma_5 + \mu_3 \sigma_3 (\sigma_1+\sigma_2) \, .
\ee
It can be analyzed as three separated sectors.

The first one is made by the
fields $\sigma_1,\sigma_2,\sigma_3$ and behave exactly as in ISS.
The second one is made by the fields $\sigma_4,\sigma_5$.
Here
once again the parameter in the whole theory
associated to the node set to zero ($\mu_4$)
is crucial for the stability of the vacuum.
In fact if $\mu_4=0$ the directions
 $\sigma_4$ and $ \sigma_5$ would result massless at tree level.

The third sector is made by the other fields
and it is stabilized at tree level taking into account the
$D$ term contributions to the scalar potential
for the gauge groups $U(N_5)$ and $U(N_6)$.

Another important fact to be stressed is that in this case we are
not obliged to switch on the deformation $\mu_6$.

\subsection{$L^{1n1}$}
\label{luciano4}
The analysis made in the last example can be extended to the
gauge theory obtained from the $L^{1n1}$ quiver
as explained in the text.
The vacuum is chosen as a natural generalization of the previous examples,
and the fluctuation superpotential has the same structure.
The non supersymmetric sector is the same than in ISS. The supersymmetric
sector is decoupled in three different parts as in the last subsection.
The tree level flat directions are stabilized provided
the deformation associated with the node set to zero and
to the first and the last nodes
are switched on.

Another requirement
for stabilizing  the flat directions in the
$L^{1n1}$ theories with $n>3$
is to take into account
the tree level $D$-term potential
of some of the flavour groups. Note that
for these nodes we need to consider also the $U(1)$
contribution to the $D$-term potential
of the $U(n)$ groups.
Otherwise, if the $U(1)$'s decouple,
some flat directions due to the trace
part of the fundamental fields can remain
in the one loop spectrum.
It would be interesting to
explore their two loop behaviour.

\subsection{Three nodes with two adjoint fields}
\label{luciano5}
\begin{figure}[h!!!]
\begin{center}
  \includegraphics[width=4cm]{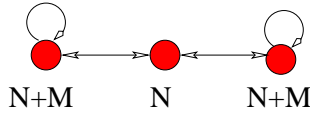}
\caption{The quiver for the $L^{222}$ theory with a node set to zero}
\label{Appendix4}
\end{center}
\end{figure}

We analyze the quiver gauge theory
of figure \ref{Appendix4}
with superpotential
\be
W=X_{11} q_{12} q_{21}-\mu_1^2 X_{11} - h q_{12} q_{23} q_{32} q_{21}+h m_1^2 q_{12}q_{21}
+ h m_3^2 q_{32} q_{23}+X_{33} q_{32} q_{23} - \mu_3^2 X_{33} \, .
\ee
We keep the more general situation
arising from the geometries analyzed in the paper.
That is the adjoint fields have linear terms and
the quarks have masses generally different from the
deformations of the adjoint field.
The choice of the ranks for the gauge groups is
\be
N_1=N_3=N+M \qquad N_2=N
\ee
with $2M>N$ and so we are guaranteed
that the second node is infrared free.
We consider this infrared free group as
the most strongly coupled.

Solving the equation of motion and
expanding around the tree level minimum
we have
\bea
&&q_{12}=
\left(
\begin{array}{c}
\mu_1+ \sigma_1 \\
\phi_1
\end{array}
\right)
\quad
q_{21}
\left(
\begin{array}{c c}
\mu_1+\sigma_2 & \phi_2
\end{array}
\right)
\quad
X_{11}=
\left(
\begin{array}{c c}
h (\mu_3^2 -  m_1^2 )+\sigma_3 & \phi_3 \\
\phi_4 & \chi_1
\end{array}
\right)
\non
\\
&&
q_{32}=
\left(
\begin{array}{c}
\mu_3+ \sigma_5 \\
\phi_5
\end{array}
\right)
\quad
q_{23}
\left(
\begin{array}{c c}
\mu_3+\sigma_6 & \phi_6
\end{array}
\right)
\quad
X_{33}=
\left(
\begin{array}{c c}
 h (\mu_1^2- m_3^2) +\sigma_7 & \phi_7 \\
\phi_8 & \chi_2
\end{array}
\right) \non 
\eea
where $\chi_1$ and $\chi_2$ are the pseudomoduli.
The superpotential for the supersymmetry breaking sector is
\bea
W&=&\chi_1 \phi_1 \phi_2-\mu_1^2 \chi_1+\mu_1 (\phi_1 \phi_4+\phi_2 \phi_3)-
h (\mu_3^2-m_1^2) \phi_1 \phi_2+ \non \\
&+&\chi_2 \phi_5 \phi_6-\mu_3^2 \chi_2+\mu_3 (\phi_5 \phi_8+\phi_6 \phi_7)-
h (\mu_1^2-m_3^2) \phi_5 \phi_6
\eea
and it consists in two O'Raifeartaigh like models
after shifting the pseudomoduli as
$\chi_1'= \chi_1 - h (\mu_3^2-m_1^2)$ and $\chi_2' =\chi_2- h (\mu_1^2-m_3^2)$.
Hence the pseudomoduli are stabilized at $\chi_1'=\chi_2'=0$
such that the non supersymmetric vacuum at quantum level
is where the mesons $X_{11}$ and $X_{33}$ are
proportional to the identity.

\section{Stability and UV completion}
\label{UVcomplete}
In this Appendix we discuss the issue of UV completion.
A related problem concerns the
unstable directions that can arise when we set some node to
zero.
The most natural UV completion to the IR theories
analyzed in this paper seems to describe them as
the last step of a duality cascade.
If this is the case there could be
potentially dangerous baryonic flat directions,
due to the breaking of the baryonic symmetry. It occurs
if we choose the baryonic branch after the confinement of some of the
gauge groups.
For supersymmetry, the Goldstone boson
associated to the breaking of baryonic symmetry
fits in a chiral supermultiplet
containing another scalar particle that is
not
protected by any symmetry. This particle is a pseudogoldstone
and signals a dangerous flat direction.

This scalar mode is decoupled at one loop and studying the
stability of this direction remains an open problem.
This was the case in \cite{Franco:2006es,Argurio:2006ny,Malyshev:2007yb}.
A possible solution is the gauging of the
baryonic symmetry. The resulting $D$-term potential
lift these dangerous directions.
Another possible way out, as noticed in \cite{Argurio:2006ny},
is to consider non canonical terms in the kahler potential.
We comment on this problem and discuss it in a simple example,
the $L^{444}$ theory.

\begin{figure}[h!!!]
\begin{center}
\includegraphics[width=8cm]{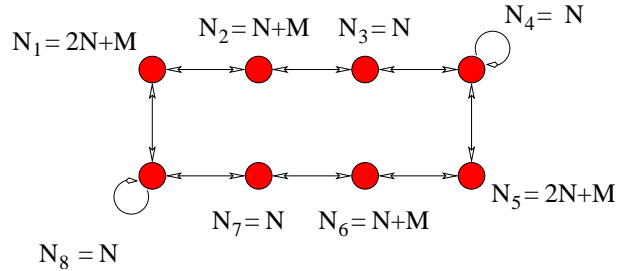}
\caption{The $L^{444}$ theory
which gives metastable vacua after the
confinement of the nodes $N_1$ and $N_5$}
\label{L444}
\end{center}
\end{figure}

We consider
the quiver in figure \ref{L444} and
we study its low energy dynamics.
Tuning the scales such that the first and the fifth node are the
more strongly coupled gauge groups, we can describe the low energy
with gauge singlets for these groups as
\bea
W=&&
q_{23}q_{34}q_{43}q_{32}
-q_{34}X_{44}q_{43}
+M_{44}X_{44}
-M_{46}M_{64}
+q_{76}M_{66}q_{67}
-\xi_1 M_{66} + \xi_1X_{44}\nonumber \\
-&&q_{67}q_{78}q_{87}q_{76}
+q_{78}X_{88}q_{87}
-M_{88}X_{88}
+M_{82}M_{28}
-q_{32}M_{22}q_{23}
+\xi_2 M_{22}- \xi_2X_{88} \, .\nonumber 
\eea
We observe that for the first and the fifth nodes the
number of flavour coincides with the number of colors.
Hence we have to impose the following quantum
constraint on the moduli space

\bea
&&\det
\left(
\begin{array}{cc}
M_{44}&M_{46}\\
M_{64}&M_{66}
\end{array}
\right)
-b_1 \tilde b_1
-\Lambda_1^{4N+2M}=0
\nonumber \\
&&
\det
\left(
\begin{array}{cc}
M_{22}&M_{28}\\
M_{82}&M_{88}
\end{array}
\right)
-b_2 \tilde b_2
-\Lambda_2^{4N+2M}=0 \, .
\eea
Choosing the baryonic branch, we have $b_i \tilde b_i = \Lambda_i^{4N+2M}$,
which breaks the baryonic symmetries.
If we integrate out the massive mesons we
obtain the low energy theory corresponding to set the nodes $N_1$
and $N_5$ to zero
\bea\label{SDUAL}
W=&& q_{76}M_{66}q_{67} -\xi_1 M_{66}
-q_{67}q_{78}q_{87}q_{76}\nonumber \\
-&&
q_{32}M_{22}q_{23}
+\xi_2 M_{22}
+q_{23}q_{34}q_{43}q_{32} \, .
\eea
This superpotential corresponds to two decoupled copies
of theories
obtained from $L^{131}$ setting to zero a node with an adjoint field,
and where we set the two deformations to have the same value but opposite
sign. This implies that there is not a mass term for the quarks.
The two theories have metastable vacua, as shown in section
\ref{Lunotreuno}.

As mentioned, the problem here is that the breaking of the global
baryonic symmetry gives rise to a Goldstone boson and to a pseudoflat
direction, which is not protected by any global symmetry.  This
direction does not receive any one loop contribution by the CW
effective potential, and can get tachyonic at higher loops. The
possible way out to this source of instability is that we are dealing
with a compactified theory. This implies that the baryonic symmetry is
gauged, and this gauging gives origin to a positive squared mass term
for the pseudoflat direction.


\begin{thebibliography}{99}



\bibitem{Intriligator:2006dd}
  K.~Intriligator, N.~Seiberg and D.~Shih,
  JHEP {\bf 0604}, 021 (2006)
  [arXiv:hep-th/0602239].


\bibitem{Franco:2006es}
  S.~Franco and A.~M.~..~Uranga,
  JHEP {\bf 0606}, 031 (2006)
  [arXiv:hep-th/0604136].


  H.~Ooguri and Y.~Ookouchi,
  Nucl.\ Phys.\  B {\bf 755}, 239 (2006)
  [arXiv:hep-th/0606061].

  A.~Amariti, L.~Girardello and A.~Mariotti,
  JHEP {\bf 0612}, 058 (2006)
  [arXiv:hep-th/0608063].

  M.~Eto, K.~Hashimoto and S.~Terashima,
  JHEP {\bf 0703}, 061 (2007)
  [arXiv:hep-th/0610042].

  E.~Dudas, C.~Papineau and S.~Pokorski,
  JHEP {\bf 0702}, 028 (2007)
  [arXiv:hep-th/0610297].


  L.~Anguelova, R.~Ricci and S.~Thomas,
  Phys.\ Rev.\  D {\bf 77}, 025036 (2008)
  [arXiv:hep-th/0702168].



  S.~Hirano,
  JHEP {\bf 0705}, 064 (2007)
  [arXiv:hep-th/0703272].

  K.~Intriligator, N.~Seiberg and D.~Shih,
  JHEP {\bf 0707}, 017 (2007)
  [arXiv:hep-th/0703281].

  I.~Garcia-Etxebarria, F.~Saad and A.~M.~Uranga,
  JHEP {\bf 0705}, 047 (2007)
  [arXiv:0704.0166 [hep-th]].

  C.~Angelantonj and E.~Dudas,
  Phys.\ Lett.\  B {\bf 651}, 239 (2007)
  [arXiv:0704.2553 [hep-th]].

  H.~Ooguri, Y.~Ookouchi and C.~S.~Park,
  arXiv:0704.3613 [hep-th].
%

  E.~Dudas, J.~Mourad and F.~Nitti,
  JHEP {\bf 0708}, 057 (2007)
  [arXiv:0706.1269 [hep-th]].


%
  R.~Essig, K.~Sinha and G.~Torroba,
  JHEP {\bf 0709}, 032 (2007)
  [arXiv:0707.0007 [hep-th]].



  H.~Abe, T.~Higaki and T.~Kobayashi,
  Phys.\ Rev.\  D {\bf 76} (2007) 105003
  [arXiv:0707.2671 [hep-th]].

  R.~Tatar and B.~Wetenhall,
  Phys.\ Rev.\  D {\bf 76}, 126011 (2007)
  [arXiv:0707.2712 [hep-th]].

  H.~Abe, T.~Kobayashi and Y.~Omura,
  JHEP {\bf 0711}, 044 (2007)
  [arXiv:0708.3148 [hep-th]].

  L.~Anguelova and V.~Calo,
  arXiv:0708.4159 [hep-th].

  Y.~Nakayama, M.~Yamazaki and T.~T.~Yanagida,
  arXiv:0710.0001 [hep-th].

  A.~Giveon and D.~Kutasov,
  arXiv:0710.0894 [hep-th].

  J.~Marsano, H.~Ooguri, Y.~Ookouchi and C.~S.~Park,
  arXiv:0712.3305 [hep-th].

  C.~Papineau,
  arXiv:0802.1861 [hep-th].

  M.~Arai, C.~Montonen, N.~Okada and S.~Sasaki,
  JHEP {\bf 0803} (2008) 004
  [arXiv:0712.4252 [hep-th]].

  M.~Arai, C.~Montonen, N.~Okada and S.~Sasaki,
  Phys.\ Rev.\  D {\bf 76} (2007) 125009
  [arXiv:0708.0668 [hep-th]].




\bibitem{Franco:2006ht}
  S.~Franco, I.~Garcia-Etxebarria and A.~M.~Uranga,
  JHEP {\bf 0701}, 085 (2007)
  [arXiv:hep-th/0607218].

  H.~Ooguri and Y.~Ookouchi,
  Phys.\ Lett.\  B {\bf 641} (2006) 323
  [arXiv:hep-th/0607183].

  I.~Bena, E.~Gorbatov, S.~Hellerman, N.~Seiberg and D.~Shih,
  JHEP {\bf 0611}, 088 (2006)
  [arXiv:hep-th/0608157].

  C.~Ahn,
  Class.\ Quant.\ Grav.\  {\bf 24}, 1359 (2007)
  [arXiv:hep-th/0608160].

  R.~Tatar and B.~Wetenhall,
  JHEP {\bf 0702}, 020 (2007)
  [arXiv:hep-th/0611303].

  A.~Giveon and D.~Kutasov,
  Nucl.\ Phys.\  B {\bf 778}, 129 (2007)
  [arXiv:hep-th/0703135].

  A.~Giveon and D.~Kutasov,
  arXiv:0710.1833 [hep-th].

  R.~Tatar and B.~Wetenhall,
  arXiv:0711.2534 [hep-th].



\bibitem{Argurio:2006ny}
  R.~Argurio, M.~Bertolini, S.~Franco and S.~Kachru,
  JHEP {\bf 0701}, 083 (2007)
  [arXiv:hep-th/0610212].

\bibitem{Aganagic:2006ex}
  M.~Aganagic, C.~Beem, J.~Seo and C.~Vafa,
  Nucl.\ Phys.\  B {\bf 789}, 382 (2008)
  [arXiv:hep-th/0610249].

\bibitem{Heckman:2007wk}
  J.~J.~Heckman, J.~Seo and C.~Vafa,
  JHEP {\bf 0707}, 073 (2007)
  [arXiv:hep-th/0702077].

\bibitem{Argurio:2007qk}
  R.~Argurio, M.~Bertolini, S.~Franco and S.~Kachru,
  JHEP {\bf 0706}, 017 (2007)
  [arXiv:hep-th/0703236].

\bibitem{Marsano:2007fe}
  J.~Marsano, K.~Papadodimas and M.~Shigemori,
  Nucl.\ Phys.\  B {\bf 789}, 294 (2008)
  [arXiv:0705.0983 [hep-th]].

\bibitem{Malyshev:2007yb}
  D.~Malyshev,
  arXiv:0705.3281 [hep-th].


\bibitem{Heckman:2007ub}
  J.~J.~Heckman and C.~Vafa,
  arXiv:0707.4011 [hep-th].


\bibitem{Aharony:2007db}
  O.~Aharony, S.~Kachru and E.~Silverstein,
  Phys.\ Rev.\  D {\bf 76}, 126009 (2007)
  [arXiv:0708.0493 [hep-th]].

\bibitem{Aganagic:2007kd}
  M.~Aganagic, C.~Beem and B.~Freivogel,
  Nucl.\ Phys.\  B {\bf 795}, 291 (2008)
  [arXiv:0708.0596 [hep-th]].

 \bibitem{DeWolfe:2008zy}
   O.~DeWolfe, S.~Kachru and M.~Mulligan,
  arXiv:0801.1520 [hep-th].

\bibitem{Marsano:2008ts}
  J.~Marsano, K.~Papadodimas and M.~Shigemori,
  arXiv:0801.2154 [hep-th].

\bibitem{Buican:2007is}
  M.~Buican, D.~Malyshev and H.~Verlinde,
  arXiv:0710.5519 [hep-th].


\bibitem{Aganagic:2007py}
  M.~Aganagic, C.~Beem and S.~Kachru,
  arXiv:0709.4277 [hep-th].



\bibitem{Altman}
K.~Altmann,
alg-geom/9403004v2


\bibitem{Franco:2005zu}
  S.~Franco, A.~Hanany, F.~Saad and A.~M.~Uranga,
  JHEP {\bf 0601}, 011 (2006)
  [arXiv:hep-th/0505040].


\bibitem{Douglas:1997de}
  M.~R.~Douglas, B.~R.~Greene and D.~R.~Morrison,
  Nucl.\ Phys.\  B {\bf 506} (1997) 84
  [arXiv:hep-th/9704151].D.~R.~Morrison and M.~R.~Plesser,
  Adv.\ Theor.\ Math.\ Phys.\  {\bf 3}, 1 (1999)
  [arXiv:hep-th/9810201].


\bibitem{Klebanov:2000hb}
  I.~R.~Klebanov and M.~J.~Strassler,
  JHEP {\bf 0008} (2000) 052
  [arXiv:hep-th/0007191].



\bibitem{Franco:2005fd}
 S.~Franco, A.~Hanany and A.~M.~Uranga,
  JHEP {\bf 0509} (2005) 028
  [arXiv:hep-th/0502113].


\bibitem{Cvetic:2005ft}
  M.~Cvetic, H.~Lu, D.~N.~Page and C.~N.~Pope,
  Phys.\ Rev.\ Lett.\  {\bf 95}, 071101 (2005)
  [arXiv:hep-th/0504225].


\bibitem{Martelli:2005wy}
  D.~Martelli and J.~Sparks,
  Phys.\ Lett.\  B {\bf 621} (2005) 208
  [arXiv:hep-th/0505027].


\bibitem{Benvenuti:2005ja}
  S.~Benvenuti and M.~Kruczenski,
  JHEP {\bf 0604} (2006) 033
  [arXiv:hep-th/0505206].

\bibitem{Butti:2005sw}
  A.~Butti, D.~Forcella and A.~Zaffaroni,
  JHEP {\bf 0509} (2005) 018
  [arXiv:hep-th/0505220].


\bibitem{Franco:2005sm}
  S.~Franco, A.~Hanany, D.~Martelli, J.~Sparks, D.~Vegh and B.~Wecht,
  JHEP {\bf 0601} (2006) 128
  [arXiv:hep-th/0505211].

\bibitem{Amariti:2007am}
  A.~Amariti, L.~Girardello and A.~Mariotti,
  JHEP {\bf 0710}, 017 (2007)
  [arXiv:0706.3151 [hep-th]].


\bibitem{GarciaEtxebarria:2006rw}
  I.~Garcia-Etxebarria, F.~Saad and A.~M.~Uranga,
  JHEP {\bf 0608} (2006) 069
  [arXiv:hep-th/0605166].


\bibitem{laba}
  M.~Cvetic, H.~Lu, D.~N.~Page and C.~N.~Pope,
  Phys.\ Rev.\ Lett.\  {\bf 95} (2005) 071101
  [arXiv:hep-th/0504225].
  D.~Martelli and J.~Sparks,
  Phys.\ Lett.\  B {\bf 621} (2005) 208
  [arXiv:hep-th/0505027].


\bibitem{Seiberg:1994pq}
  N.~Seiberg,
  Nucl.\ Phys.\  B {\bf 435}, 129 (1995)
  [arXiv:hep-th/9411149].


\bibitem{Intriligator:2005aw}
  K.~Intriligator and N.~Seiberg,
  JHEP {\bf 0602} (2006) 031
  [arXiv:hep-th/0512347].



\bibitem{Brini:2006ej}
  A.~Brini and D.~Forcella,
  JHEP {\bf 0606}, 050 (2006)
  [arXiv:hep-th/0603245].


\bibitem{Dine:1981gu}
  M.~Dine and W.~Fischler,
  Phys.\ Lett.\  B {\bf 110}, 227 (1982).

  M.~Dine and W.~Fischler,
  Nucl.\ Phys.\  B {\bf 204}, 346 (1982).

  G.~F.~Giudice and R.~Rattazzi,
  Phys.\ Rept.\  {\bf 322}, 419 (1999)
  [arXiv:hep-ph/9801271].

\bibitem{Dine:2006xt}
  M.~Dine and J.~Mason,
  Phys.\ Rev.\  D {\bf 77}, 016005 (2008)
  [arXiv:hep-ph/0611312].


  R.~Kitano, H.~Ooguri and Y.~Ookouchi,
  Phys.\ Rev.\  D {\bf 75}, 045022 (2007)
  [arXiv:hep-ph/0612139].

  C.~Csaki, Y.~Shirman and J.~Terning,
  JHEP {\bf 0705}, 099 (2007)
  [arXiv:hep-ph/0612241].

  O.~Aharony and N.~Seiberg,
  JHEP {\bf 0702}, 054 (2007)
  [arXiv:hep-ph/0612308].

  S.~A.~Abel and V.~V.~Khoze,
  arXiv:hep-ph/0701069.

  A.~Amariti, L.~Girardello and A.~Mariotti,
  Fortsch.\ Phys.\  {\bf 55}, 627 (2007)
  [arXiv:hep-th/0701121].

  H.~Murayama and Y.~Nomura,
  Phys.\ Rev.\  D {\bf 75}, 095011 (2007)
  [arXiv:hep-ph/0701231].


  D.~Shih,
  arXiv:hep-th/0703196.

  T.~Kawano, H.~Ooguri and Y.~Ookouchi,
  Phys.\ Lett.\  B {\bf 652}, 40 (2007)
  [arXiv:0704.1085 [hep-th]].


  J.~E.~Kim,
  Phys.\ Lett.\  B {\bf 651}, 407 (2007)
  [arXiv:0706.0293 [hep-ph]].

  H.~Y.~Cho and J.~C.~Park,
  JHEP {\bf 0709}, 122 (2007)
  [arXiv:0707.0716 [hep-ph]].

  S.~Abel, C.~Durnford, J.~Jaeckel and V.~V.~Khoze,
  arXiv:0707.2958 [hep-ph].


%
  J.~E.~Kim,
  Phys.\ Lett.\  B {\bf 656}, 207 (2007)
  [arXiv:0707.3292 [hep-ph]].

  N.~Haba and N.~Maru,
  Phys.\ Rev.\  D {\bf 76} (2007) 115019
  [arXiv:0709.2945 [hep-ph]].

  N.~Haba,
  arXiv:0802.1758 [hep-ph].



\bibitem{Hanany:2006nm}
  A.~Hanany, C.~P.~Herzog and D.~Vegh,
  JHEP {\bf 0607} (2006) 001
  [arXiv:hep-th/0602041].


\bibitem{Butti:2006hc}
  A.~Butti,
  JHEP {\bf 0610} (2006) 080
  [arXiv:hep-th/0603253].


\bibitem{Butti:2006nk}
  A.~Butti, D.~Forcella and A.~Zaffaroni,
  JHEP {\bf 0702} (2007) 081
  [arXiv:hep-th/0607147].

\bibitem{Benvenuti:2006qr}
  S.~Benvenuti, B.~Feng, A.~Hanany and Y.~H.~He,
  JHEP {\bf 0711} (2007) 050
  [arXiv:hep-th/0608050].

\bibitem{Butti:2006au}
  A.~Butti, D.~Forcella and A.~Zaffaroni,
  JHEP {\bf 0706} (2007) 069
  [arXiv:hep-th/0611229].



\bibitem{Forste:2006zc}
  S.~Forste,
  Phys.\ Lett.\  B {\bf 642}, 142 (2006)
  [arXiv:hep-th/0608036].

\end{thebibliography}
\end{document}